\documentclass[preprint,showpacs,preprintnumbers,amsmath,amssymb,superscriptaddress]{revtex4}
\usepackage{amssymb}
\usepackage{amsmath}
\usepackage{graphicx}% Include figure files
\usepackage{dcolumn}% Align table columns on decimal point
\usepackage{bm}% bold math
\usepackage{times}
\usepackage{color}

\begin{document}
%\preprint{}
%
%##############################################################################################################
%\title{Dielectric and Impedance Studies of Y$_2$NiMnO$_6$ Across the Magnetic Transition}
%\title{Non-Griffith's-like Behavior and Deceptive Polarization in Y$_2$NiMnO$_6$ Double Perovskite}
\title{Ferromagnetism and the Effect of Free Charge Carriers on Electric Polarization in Y$_2$NiMnO$_6$ Double Perovskite}
%##############################################################################################################
%
\author{Hariharan Nhalil}
\email{hariharan.nhalil@gmail.com}
%\thanks{}
\affiliation{Department of Physics, Indian Institute of Science, Bangalore 560012, India}
\author{Harikrishnan S. Nair}
\affiliation{Highly Correlated Matter Research Group, Physics Department, University of Johannesburg, P. O. Box 524, Auckland Park 2006, South Africa}
\author{C. M. N. Kumar}
\affiliation{J\"{u}lich Centre for Neutron Science (JCNS), Forschungszentrum J\"{u}lich GmbH, Outstation at SNS, Oak Ridge National Laboratory, Oak Ridge, Tennessee 37831, United States}
\affiliation{Chemical and Engineering Materials Division, Oak Ridge National Laboratory, Oak Ridge, Tennessee 37831, United States}
%
%\author{Ashfia Huq}
%\affiliation{Spallation Neutron Source, Oak Ridge National Laboratory, Oak Ridge, Tennessee 37831-6473, USA}
%
\author{Andr\'{e} M. Strydom}
\affiliation{Highly Correlated Matter Research Group, Physics Department, University of Johannesburg, P. O. Box 524, Auckland Park 2006, South Africa}
%\affiliation{Max Planck Institute for Chemical Physics of Solids (MPICPfS), N\"{o}thnitzerstra{\ss}e 40, 01187 Dresden, Germany}
%
\author{Suja Elizabeth}
\affiliation{Department of Physics, Indian Institute of Science, Bangalore 560012, India}
%\noaffiliation is required (may also be used with the \author command).
\date{\today}
%
%#############################################################################################################
%
\begin{abstract}
	\indent 
	The double perovskite  Y$_2$NiMnO$_6$ displays ferromagnetic transition at $T_c \approx$ 81~K. The ferromagnetic order at low temperature is confirmed by the saturation value of magnetization ($M_s$) and also, validated by the refined ordered magnetic moment values extracted from neutron powder diffraction data at 10~K. This way, the dominant Mn$^{4+}$ and Ni$^{2+}$ cationic ordering is confirmed. The cation-ordered $P2_1/n$ nuclear structure is revealed by neutron powder diffraction studies at 300 and 10~K. Analysis of frequency dependent dielectric constant and equivalent circuit analysis of impedance data takes into account the bulk contribution to total dielectric constant. This reveals an anomaly which coincides with the ferromagnetic transition temperature ($T_c$). Pyrocurrent measurements register a current flow with onset near $T_c$ and a peak at 57~K that shifts with temperature ramp rate. The extrinsic nature of the observed pyrocurrent is established by employing a special protocol measurement. It is realized that the origin is due to re-orientation of electric dipoles created by the free charge carriers and not by spontaneous electric polarization at variance with recently reported magnetism-driven ferroelectricity in this material. 
\end{abstract}
%Use showkeys class option if keyword
\maketitle
%###############################################################################################################
%
%
%
\section{Introduction}
\label{intro}
\indent 
\indent 
The double perovskite (DP) family of compounds with chemical formula $R_2M'M''$O$_6$ ($R$ = rare earth, $M'/M''$ = transition metal) became an important research topic in the search for multifunctional materials by virtue of the wide variety of material properties shown by them, such as ferromagnetism,  magnetocapacitance (MC), magnetoresistance (MR) and spin-phonon coupling.\cite{rogado2005magnetocapacitance,nair2011griffiths,guo2013near,macedo2013spin} Magnetic behavior of DPs is complex as a result of various exchange interactions between ions with different valencies ($M'^{3+}$, $M'^{4+}$, $M''^{3+}$, $M''^{2+}$, etc.) mediated through the intervening oxygen. These  valencies are stabilized in DPs due to the presence of "antisite" disorder where $M'$ and $M''$ cations interchange their respective crystallographic positions.\cite{anderson1993b} The ground state magnetic properties of double perovskites are determined by the degree of cationic ordering of the transition metal ions, $M'$ and $M''$. In ordered DPs $i.e.,$ where the $M'$ and $M''$ ions are ordered crystallographically, $M'^{4+}$ and $M''^{2+}$ layers alternate periodically.\cite{anderson1993b} Ordered $R_2$NiMnO$_6$  DPs are ferromagnetic (FM) due to the Mn$^{4+}$--O--Ni$^{2+}$ superexchange interaction conforming to Goodenough-Kanamori rules.\cite{goodenough1955theory} The inevitable presence of "antisite" disorder can introduce additional antiferromagnetic (AFM) interactions across the exchange pathways such as Mn$^{4+}$--O--Mn$^{4+}$ and Ni$^{2+}$--O--Ni$^{2+}$.\cite{zhou2010nature} \\
\indent
A recent excitement in the field of DPs  has risen from the prospect of realizing multiferroics. According to density functional theory calculations, $R_2$NiMnO$_6$ series of compounds with small rare earth ions are predicted to be multiferroic from the nature of $E^*$-type ($\uparrow\uparrow\downarrow\downarrow$ ) magnetic ordering which breaks the inversion symmetry and generates spontaneous electric polarization.\cite{kumar2010theoretical} More recently, experimental evidences for multiferroicity have been reported in Lu$_2$CoMnO$_6$ \cite{yanez2011multiferroic} and Y$_2$CoMnO$_6$.\cite{sharma2013magnetism} In these compounds, the ferroelectric transition coincides with the magnetic transition. The $E^*$-type spin structure, $\uparrow\uparrow\downarrow\downarrow$ of  Mn$^{4+}$ and  Co$^{2+}$ along $c$-axis was projected as the origin of ferroelectricity in Lu$_2$CoMnO$_6$.\cite{yanez2011multiferroic} 
\\
\indent
Y$_2$NiMnO$_6$ (YNMO), the title compound of this paper, was reported to show a FM transition at $T_c\approx$~85~K  with claims of magnetically-driven ferroelectricity in the material.\cite{su2015magnetism} The low temperature dielectric properties of YNMO, however, did not present any major anomaly corresponding to the magnetic transition temperature ($T_c$).\cite{kakarla2014dielectric,mouallem2004nuclear} Even though direct evidence of magneto-dielectric coupling does not exist, we can not rule out the possibility of such a phenomenon based on the measurement of dielectric constant alone. Equivalent circuit (EC) analysis of dielectric and impedance data may shed more light on this, similar to the case of LaCoO$_3$ and BiMnO$_3$ where, despite the absence of dielectric anomaly at T$_c$ magneto-dielectric coupling was established.\cite{Schmidt2007,Schmidt2012,schmidt2009dielectric} In this paper, we combine the results in Y$_2$NiMnO$_6$ from a suite of measurements comprising of magnetometry, dielectric and impedance measurements and neutron powder diffraction so as to examine the magnetic and dielectric behavior closely.
\section{Experimental Details}
\label{exp}
\indent
Polycrystalline YNMO compound was prepared by conventional solid state synthesis route similar to that used in the case of other DPs.\cite{nair2011griffiths,nair2014magnetizationsteps} High purity precursors Y$_2$O$_3$, MnO$_2$ and NiO (4$N$, Sigma Aldrich) were ground thoroughly and heated at 1300$^{\circ}$ C for 24~h for several times with intermediate grinding. Powder X-ray diffractograms were obtained to check the phase formation. Preliminary structural refinements indicated monoclinic $P2_1/n$ symmetry similar to other DPs.\cite{nair2014magnetizationsteps} Magnetization data was obtained in a Magnetic Property Measurement System (Quantum Design Inc.). In order to make dielectric, impedance and pyrocurrent measurements, the powder sample was pressed into pellets. Electrodes were formed on both sides of the pellet by applying thick silver paste. Dielectric and impedance response was recorded in the temperature range 15-300~K using a HP-4294 high precision impedance analyzer with applied AC voltage of 500~mV. Pyrocurrents were measured in a Keithely-6514 electrometer. To investigate nuclear and magnetic structure, time-of-flight (TOF) neutron powder diffraction (NPD) experiments were performed at the Spallation Neutron Source (SNS) at Oak Ridge National Laboratory using high resolution neutron powder diffractometer POWGEN.\cite{AHuq_1ZK2011AThirdGeneration} Measurements were made on a 3.8 gm sample in a vanadium can (8 mm diameter) at  10 and 300~K. At each temperature, two different central wavelengths (CWL, 1.066 and 2.665~{\AA}) were used to obtain the patterns. The first CWL covers the $d$-spacing from 0.3 to 4.6~{\AA} and probes the nuclear structure. The second CWL covers higher $d$-spacing from 0.7 to 5.5~{\AA} which provides insight into magnetic contributions. Crystal and magnetic structure refinements were carried out using the NPD data in conjunction with Rietveld refinement \cite{rietveld} program FullProf.\cite{fullprof}
\section{Results}
\subsection{Ferromagnetism and Non-Griffiths-like features}
\label{griffiths}
\begin{figure}[!t]
	\centering
	\includegraphics[scale=0.35]{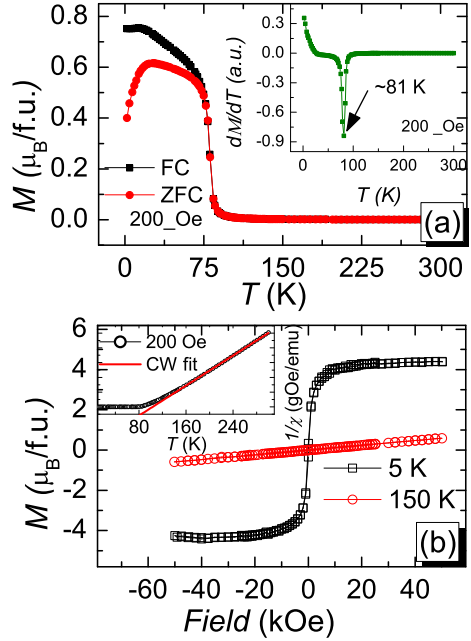}
	\caption{\label{fig1_mag}(colour online) (a) Temperature dependence of ZFC and FC magnetization in the warming cycle measured at 200~Oe. Inset of (a) exhibits the derivative plot of magnetization, $dM/dT$, to elucidate the transition temperature, $T_c$ = 81~K. (b) Isothermal magnetization plots at 5~K suggests that YNMO is a soft ferromagnet. Inset of (b) shows the 1/$\chi$ (T) plot and Curie-Weiss fit at 200 Oe field. Slight upward deviation from the fit is clearly seen.}
\end{figure}

\indent
The magnetic response of YNMO presented in Fig.~\ref{fig1_mag} relates well with the overall FM features. Zero-field cooled (ZFC) and  field-cooled (FC)  magnetization data at 200 Oe in the range 5--300~K are shown in Fig.~\ref{fig1_mag} (a). A sharp FM transition is seen at $T_c$ = ~81~K. $T_c$ is accurately estimated from the derivative plot of magnetization data (inset of Fig.~\ref{fig1_mag} (a)). The ZFC curve has a downturn at low temperature (below 10~K) seen clearly in low- field plots. At high field, say 20~kOe (data not shown), the downturn disappears and the ZFC curve overlaps with the FC curve. Here, the maximum magnetic moment at 5~K is $\approx$ 4.2~$\mu_\mathrm B$/f.u.  It is noted that such a downturn occurs in R$_2$NiMnO$_6$ systems irrespective of whether $R$ ion is magnetic or not. This rules out the role of R and Ni/Mn sub-lattice effects.\cite{kakarla2014dielectric} Also, through AC susceptibility studies, absence of spin-glass behavior is reported in YNMO.\cite{maiti2012magnetic} The low temperature downturn is supposed to originate from antiphase FM domains resulting from spatial distribution of Ni$^{2+}$/Mn$^{4+}$ ordered domains and out-of-phase ordering.\cite{asaka2007strong}
\\
%
% FIGURE 1 ######################################################################################################################
%################################################################################################################################
%
\indent
Fig.~\ref{fig1_mag} (b) gives the isothermal magnetization plots at 5~K and 150~K. Magnetization saturates at about 20~kOe. The saturation magnetization, $M_{s}$ is estimated as 4.3(1)~$\mu_\mathrm{B}$/f.u. at 5~K and 50~kOe. If we assume a mixture of Mn$^{4+}$ (3$d^{3}$; $S$ = $\frac{3}{2}$) and Ni$^{2+}$ (3$d^{8} $, $S$ = 1), spin-only moment yields 4.79~$\mu_\mathrm{B}$/f.u. In the case of Mn$^{3+}$ (3$d^{4}$, $S$ = 2) and Ni$^{3+}$ (3$d^{7}$, $S$ = $\frac{3}{2}$) combination, it gives 6.24~$\mu_\mathrm{B}$/f.u. The experimentally observed $M_{s} $ features a dominant Mn$^{4+}$--O--Ni$^{2+}$ superexchange interaction that favors ferromagnetism. Antisite disorder of Mn$^{4+}$ and Ni$^{2+}$ has a marked influence on the magnetic interactions in DPs as the Mn$^{4+}$--O--Mn$^{4+}$ and Ni$^{2+}$--O--Ni$^{2+}$ exchange paths are AFM whereas the Mn$^{4+}$--O--Ni$^{2+}$ paths are FM.\cite{goodenough1955theory,dass2003oxygen,ogale1999octahedral} A small hysteresis is observed at 5~K in YNMO which is suggestive of a soft ferromagnet. This behavior is in contrast to YNMO reported earlier \cite{su2015magnetism}, Tb$_2$NiMnO$_6$ \cite{nair2011griffiths} and Y$_2$CoMnO$_6$ \cite{nair2014magnetizationsteps} all of which are DPs but display different degrees of cationic disorder and different features in their respective magnetic hysteresis.
\\
\indent
The inverse magnetic susceptibility curve, 1/$\chi(T)$, at 200~Oe along with the Curie-Weiss (CW) fit is shown in inset of Fig.~\ref{fig1_mag} (b). The solid line gives the fit between 300 and 140~K. The analysis yields the Curie-Weiss constant, $\Theta_{CW}$ = 84~K in support of predominant FM interactions. Estimated value of effective paramagnetic moment, $\mu_\mathrm{eff}$ is 4.8(3)~$\mu_\mathrm B$/f.u., which is close to the theoretically calculated effective moment 4.79~$\mu_\mathrm{B}$/f.u. by assuming Mn$^{4+}$, 3$d ^{3}$, $S$ = $\frac{3}{2}$ and Ni$^{2+}$, 3$d^{8}$, $S$ = 1 combination. A notable aspect of the 1/$\chi(T)$ plot is the upward deviation from ideal CW description at $T\approx$ 2$T_c$. This is common to conventional ferromagnets.\cite{dagotto2001colossal,li2005ferromagnetic,he2007non} An explanation advanced is based on the formation of FM clusters well above the $T_c$.\cite{he2007non} It is also notable that the upward deviation from the CW description in YNMO is at odds with the Griffiths-like model as applied to the downward trend observed in other DPs.\cite{nair2011griffiths} Magnetic field-dependence is an important factor in this context because, the downward deviation temperature in 1/$\chi(T)$ is determined by the strength of applied field in Griffiths phase systems. 
On the other hand, the non-Griffiths-phase is characterized by the absence of field-dependence.\cite{he2007non,zhou2010nature} 
This phase derives its origin from competing AFM and FM order well above the $T_c$.\cite{zhou2011size,saber2010evolution} In  YNMO, the overlap of 200~Oe and 2~kOe data and the upward deviation from ideal CW description starts at a common temperature--140~K (fig not shown).  The occurrence of Griffiths-phase in rare earth DPs is thus strongly influenced by  antisite disorder. The relation between antisite disorder and ionic size of the rare earth emerges as an interesting detail. For example, rare earths such as La, Pr, Tb \cite{biswal_jap_115_2014dielectric,liu_jap_116_2014griffiths,nair2011griffiths} exhibit Griffiths phase-like features while Y, as in the present case, does not. 
\subsection{Nuclear and Magnetic structures}
\label{neutrons}
\begin{figure*}[!t]
	\centering
	\includegraphics[scale=0.32]{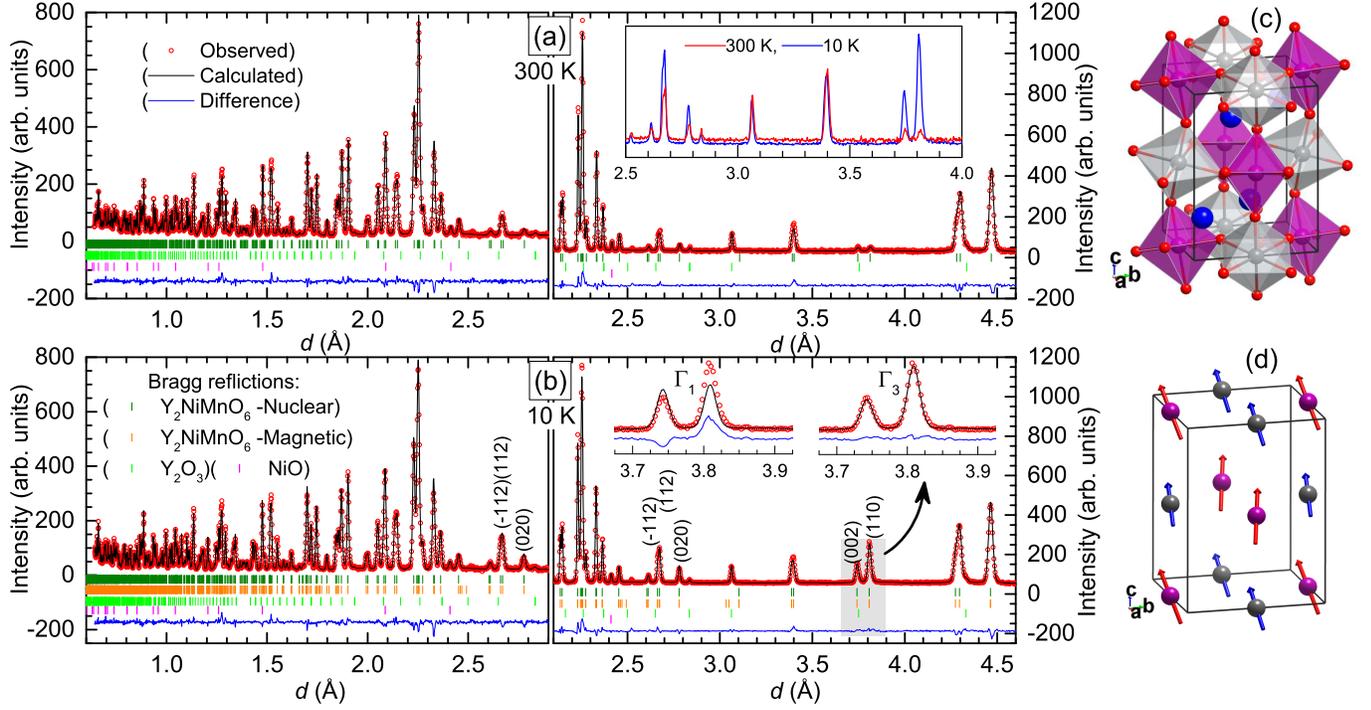}
	\caption{\label{fig_neutron} (colour online) Observed and calculated diffractions patterns and their difference at (a) 300~K and (b) 10~K. Left and right panels correspond to the data collected with central wavelength 1.066 and 2.665~{\AA} respectively. Black circles are the measured intensities and the red curve is a calculated pattern. Olive and orange lines mark the nuclear and magnetic Bragg positions respectively for YNMO. Magenta and green lines indicate the nuclear Bragg positions of Y$_2$O$_3$ and NiO respectively. Blue curve is the difference between the measured and calculated patterns. Upper inset (a) compares a part of the diffraction data collected at 300 and 10~K; the bottom insets (b) give the experimental data and the patterns calculated with $\Gamma_1$ and $\Gamma_3$ magnetic symmetry and the difference curve. (c) Nuclear and (d) magnetic structure of YNMO. Mn and Ni atoms are shown in magenta and grey colors respectively, Y and O atoms in blue and red colors respectively.  }
\end{figure*}

\indent
The nuclear and magnetic structures of YNMO at 10~K and 300~K were investigated by analyzing the time-of-flight (TOF) neutron powder diffraction data. The experimentally obtained powder diffraction data are presented in Figs.~\ref{fig_neutron} (a) and (b) at 300~K and 10~K respectively. Figs.~\ref{fig_neutron} (a) and (b) contain two panels each for the two central wavelengths  used in the TOF experiment, $viz.,$ 1.066~{\AA} and 2.665~{\AA}. At first, the 300~K data was analyzed in order to ascertain the crystal structure at room temperature and to estimate the degree of "antisite" disorder, if it was present. Rietveld refinement using two structural models -- monoclinic $P2_1/n$ and orthorhombic $Pnma$ confirmed that only the monoclinic space group allows cationic ordering of $M'$ and $M''$ cations in $R_2M'M''$O$_6$.\cite{bulljpcm_15_4927_2003} The structural model and resulting lattice parameters are collected in Table~\ref{tab1}. The analysis confirms that YNMO crystallizes in $P2_1/n$ space group with high degree ($\approx$ 97$\%$) of cationic order. Minor impurity phases of NiO ($\approx$ 3.5(3) wt$\%$) and Y$_2$O$_3$ ($\approx$ 4.0(2) wt$\%$) were identified. In the refinement cycles, "antisite" disorder (crystallographic inversion of $M'$ and $M''$ cation positions) was introduced in order to check the implications. However, no improvement in the quality of fits was found. Refinement was also carried out in the $Pnma$ space group which allows for a random distribution of $M'$ and the $M''$ cations at the $6c$ Wyckoff position. However, the quality factors obtained for the monoclinic setting were better. In conjunction with $M_s$ obtained from the magnetization measurements, we describe the nuclear structure of YNMO as cation-ordered, monoclinic $P2_1/n$, as shown schematically in Fig~\ref{fig_neutron} (c).
\\
% TABLE1 ########################################################################################################################
\begin{table}[!h]
	\caption{\label{tab1} The refined lattice parameters, atomic positions, phase fraction and magnetic moments of Y$_2$NiMnO$_6$ obtained from Rietveld refinement of the neutron diffraction data at 300~K and at 10~K. The refinements were carried out using the space group $P2_1/n$. Ni and Mn cations occupy $2c$ ($0$ $\frac{1}{2}$ $0$) and $2d$ ($\frac{1}{2}$ $0$ $0$) Wyckoff positions respectively. $M_{av}$ is the refined magnetic moment per atom and $R_w$ and $R_{wp}$ are quality factors.}
	\setlength{\tabcolsep}{15pt}
	\begin{tabular}{ccccc} \hline
		$T(K)$     &    300~K        &  10~K      	\\ \hline
		S.G.     & $P2_1/n$        & $P2_1/n$     	\\ \hline 
		$a (\AA)$        &  5.2267(3)	   & 5.2202(2)    	\\
		$b (\AA)$        &  5.5582(6)	   & 5.5534(2)	 	\\
		$c (\AA)$        &  7.4835(2)	   & 7.4745(2)	 	\\ 
		$\beta (^\circ)$ &  89.706(45) 	   & 89.697(24)	 	\\ \hline
		Y $x$            &  -0.0183(5)	   & -0.0188(6)   	\\
		$y$              &  0.0707(7)	   & 0.0716(5)	 	\\
		$z$              &  0.2497(10)	   & 0.2494(5)	 	\\
		$B_{iso}(\AA^{2})$&  0.246(5)    & 0.072(7)		\\ \hline
		O(1) $x$         &  0.1046( 18)	   & 0.1048(7)	    \\
		$y$              &  0.4650(20)	   & 0.4655(7)      \\
		$z$              &  0.2548(11)	   & 0.2559(6)	    \\
		$B_{iso}(\AA^{2})$&  0.339(3) 	   & 0.108(3)		\\ \hline
		O(2) $x$         &  0.6846(16)	   & 0.6845(8)      \\
		$y$              &  0.2966(19)	   & 0.2951(7)      \\
		$z$              &  0.0551(13)	   & 0.0542(6)	    \\
		$B_{iso} (\AA^{2})$&  0.339(3) 	   & 0.108(3)		\\ \hline
		O(3) $x$         &  0.6987(16)	   & 0.6995(8) 	    \\
		$y$              &  0.3121(17)	   & 0.3128(9) 	    \\
		$z$              &  0.4490(13)	   & 0.4487(6)	    \\
		$B_{iso}(\AA^{2})$&  0.339(3) 	   & 0.108(3)		\\ \hline
		Ni, $M_{av}(\mu_{\mathrm{B}})$     &  --	           & 1.77(10)       \\
		$B_{iso}(\AA^{2})$&  0.300(6) 	   & 0.242(7)		\\ \hline
		Mn, $M_{av}(\mu_{\mathrm{B}})$     &  --	           & 3.12(15)       \\
		$B_{iso}(\AA^{2})$&  0.600(12)    & 0.942(3)		\\ \hline 
		$R_p(\%)$          & 10.5 	       & 8.79           \\
		$R_{wp}(\%)$       & 13.5	       & 10.6           \\
		$R_{mag}(\%)$      & --	           & 3.95           \\
		$\chi^2$         & 1.02	           & 0.49	        \\ \hline
		%phase fraction (wt\%)& \textcolor{blue}{92.5(5)} & \textcolor{blue}{92.5(5)}   \\ \hline\hline
	\end{tabular}
\end{table}
%######################################################################################################################################
%
% FIGURE 3###########################################################################################################################
%
%###################################################################################################################################
%
\indent
Next, we analyzed the neutron powder diffraction data at 10~K. The inset of Fig~\ref{fig_neutron}(a) highlights the diffraction patterns at 10~K and 300~K to clarify the enhancement of intensities at low temperature due to the FM order. No superstructure peaks are observed that support the hypothesis of ferromagnetism in YNMO as earlier suggested by macroscopic magnetization data. The nuclear structure of YNMO at 10~K was refined in $P2_1/n$ space group similar to the room temperature data. Using the $\bf k$-search utility in the FullProf suite of programs, $\bf k$ = ($\bf 0 0 0$) was identified as the propagation vector of the magnetic unit cell of YNMO. Again, this is in accordance with the FM nature of the compound. From Fig~\ref{fig_neutron} (b), it can be seen that magnetic peaks ($\bf 0 0 2$) and ($\bf 1 1 0$) have enhanced intensity at 10~K. The software SARA$h$\cite{wills_sarah} was used to perform representation analysis to elucidate the magnetic structure. Consequently, two magnetic representations were identified as $\Gamma_1$ and  $\Gamma_3$. From the visual quality and reliability factors of the fit, $\Gamma_3$ was selected as the most suitable magnetic representation for YNMO. A comparison of the fits ($\Gamma_1$ and $\Gamma_3$ models) with the ($\bf 0 0 2$) and ($\bf 1 1 0$) peaks (inset of Fig~\ref{fig_neutron} (b)) clearly establishes this fact. In Fig.~\ref{fig_neutron} (d), a schematic of $\Gamma_3$ magnetic structure is presented.  First principles DFT calculations predict that the FM ground state in YNMO \cite{kumar2010theoretical} transforms to AFM $\uparrow\uparrow\downarrow\downarrow$ structure upon application of an electric field. Subsequently, it develops ferroelectric polarization of 0.75~$\mu$C/cm$^2$ along the $b$-axis. Our measurements and analysis of magnetization and magnetic structure clearly establishes FM spin structure. 
%
%
% TABLE1 ###########################################################################################################################
\begin{table}[!h]
	\caption{\label{tab2} Bond parameters of Y$_2$NiMnO$_6$ obtained after Rietveld refinement of the neutron diffraction data at 300~K and at 10~K. The theoretical estimates obtained through DFT calculations as per Ref.[\onlinecite{kumar2010theoretical}] are also shown.}
	\setlength{\tabcolsep}{7pt}
	\begin{tabular}{lllllll} \hline\hline
		Bond angle/distance  &    300~K              & 10~K             	& Theory            \\ \hline\hline
		$\angle$Ni--O1--Mn	         &  145.64(15)$^\circ$   & 145.04(15)$^\circ$   &             		\\
		$\angle$Ni--O2--Mn           &  145.54(17)$^\circ$   & 145.78(17)$^\circ$   & 146.31$^\circ$    \\
		$\angle$Ni--O3--Mn           &  146.63(16)$^\circ$   & 146.39(18)$^\circ$   & 147.0$^\circ$     \\
		Ni -- Ni           		     &  5.309(7)             & 5.26(14)        		&            		\\
		Mn -- Mn                     &  7.4812(4)            & 7.4745(14)    	  	&               	\\
		Ni -- Mn                     &  3.7783(15)           & 3.7740(11)      		&                  	\\ \hline\hline
	\end{tabular}
\end{table}
%###################################################################################################################################
%
The refined bond distances and angles of YNMO at 300~K as well as at 10~K are listed in Table~\ref{tab2}. Theses are in close agreement with the predicted values in reference [\onlinecite{kumar2010theoretical}]. Note that the bond parameters of the FM and the $E^*$-type structure which stabilizes in the presence of applied electric field are very close.
\subsection{Dielectric and Impedance Measurements} 
\label{impedance}
\begin{figure}[!t]
	\centering
	\includegraphics[scale=0.28]{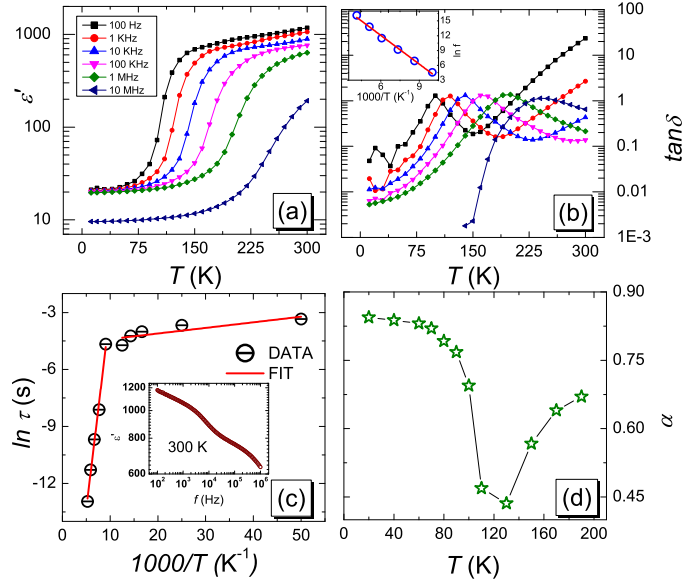}
	\caption{\label{fig3_dielectric} (colour online) The real part of dielectric constant $\epsilon'$ and the loss tangent tan$\delta$  are shown in (a) and (b) respectively. The inset of (b) depicts Arrhenius law description of the peak temperature of the loss tangent. (c) Log $\tau$ versus 1000/$T$ plot. Two regions with different activation energies above and below the $T_c$ are observed. Inset of (c) shows the $\epsilon$'(f) plot at 300 K. (d) Temperature dependence of $\alpha$ showing anomaly near $T_c$. The relaxation time, $\tau$, and the $\alpha$ values were obtained from the curve fit in $\epsilon'$ versus $f$ plot using the modified-Debye equation (Eqn.~(\ref{Debye_eqn})).}
\end{figure}

\indent
The theoretical predictions of ferroelectric polarization in YNMO were verified by measuring temperature dependent dielectric constant between 15 and 300~K at different frequencies and at the applied ac voltage of 500~mV. The measured real part of dielectric constant $\epsilon'(T)$ and loss tangent tan$\delta$ (= $\epsilon''(T)$/$\epsilon'(T)$) are shown in Figs.~\ref{fig3_dielectric} (a) and (b) respectively. Giant values of dielectric constant are seen in this compound at 300~K especially at low frequency. At low temperature, the dielectric permittivity is independent of frequency and temperature. A broad-hump like feature without frequency dependence around FM transition temperature is seen in $\epsilon'(T)$ (figure not shown). A similar anomaly is reported in YNMO by other groups claimed it as a ferroelectric transition.\cite{zhang2014hydrothermal} Interestingly, no features corresponding to this anomaly is observed in tan$\delta$ which rises doubts on the true nature of this transition. Later, we found that the broad peak shifts with change in the temperature ramp rate. For a ferroelectric transition, the peak temperature is independent of temperature ramp rate and frequency of the applied field. At high temperature, both $\epsilon'(T)$ and tan$\delta$ showed significant frequency dispersion and the peaks shifted to higher temperature regime with increasing frequency. This suggests that a thermally activated relaxation mechanism is present in YNMO.
%
% FIGURE 3 ########################################################################################################################
% ####################################################################################################################################
%
The tan$\delta$ peaks were analyzed using thermal activation model in order to elucidate the relaxation dynamics. The peak temperatures were fitted to Arrhenius law, $f$ = $f_0$ exp$(-E_a/k_BT)$; where, $f_0$ is a pre-exponential constant, $E_a$ is the activation energy and $k_B$ is the Boltzmann constant. The Arrhenius curve-fit is given in the inset of Fig.~\ref{fig3_dielectric} (b), from  which $E_a$ was estimated as 160(1)~meV and the relaxation time $\left(\frac{1}{f_0}\right)$ as 4.9(3) $\times$ 10$^{-11}$~s.
\\
\indent
Frequency-dependence of dielectric constant at different temperatures were measured in the range, 100~Hz to 1~MHz at selected temperatures. In the case of ideal Debye relaxation of non-interacting dipoles, the plot of $\epsilon'(T)$  versus  $\epsilon''(T)$ is a perfect semicircle.\cite{Cole1941} But in real systems, deviation from the ideal behavior is expected and hence  modeled after modified-Debye equation,
%
% EQUATION 1 ################################################################################################################
\begin{equation}
\epsilon^{*} = \epsilon' + i \epsilon'' = \epsilon_\infty + \frac{(\epsilon_0 - \epsilon_\infty)}{[1 + (i\omega \tau)^{(1-\alpha)}]}
\label{Debye_eqn}
\end{equation}
%############################################################################################################################
%
where, $\epsilon_0$ and $\epsilon_\infty$ are the static and high frequency dielectric constants respectively, $\omega$ is the angular frequency, $\tau$ is the mean relaxation time and $\alpha$ is a parameter which represents the distribution of relaxation times (for ideal Debye relaxation, $\alpha$ = 0). Eqn.~(\ref{Debye_eqn}) was separated into real and imaginary part of dielectric permittivity and the former fitted to the experimental data to deduce the $\alpha$-values at different temperatures. Fig.~\ref{fig3_dielectric} (c)  shows the ln($\tau$) versus 1000/$T$ plot and (d) shows the temperature dependence of $\alpha$. In Fig.~\ref{fig3_dielectric} (c), two regions with different slopes are observed, characteristics of two activation energies. The temperature at which the deviation in curve occurs coincides with the $T_c$. An anomaly is detected in the temperature dependence of $\alpha$. Arrhenius fit was used to extract the activation energies above and below $T_c$; these are, 155.0(2)~meV and 1.9(1)~meV, respectively. $E_a$ above $T_c$ matches well with the value reported for YNMO .\cite{kakarla2014dielectric} Here, hopping of charge carriers between spatially fluctuating lattice potentials has been reported as the cause of relaxation. Below the FM transition, the activation energy is two orders of magnitude lower. It is considered that the activation energy will be lower in the spin ordered state where electron hopping between ordered spins spend lower energy as compared to the hopping energy in the paramagnetic regime.\cite{Manna2014,yang2012small} This is suggestive of a correlation between the magnetic and dielectric properties in YNMO.
\\
%
% FIGURE 4 ################################################################################################################
\begin{figure}[!t]
	\centering
	\includegraphics[scale=0.28]{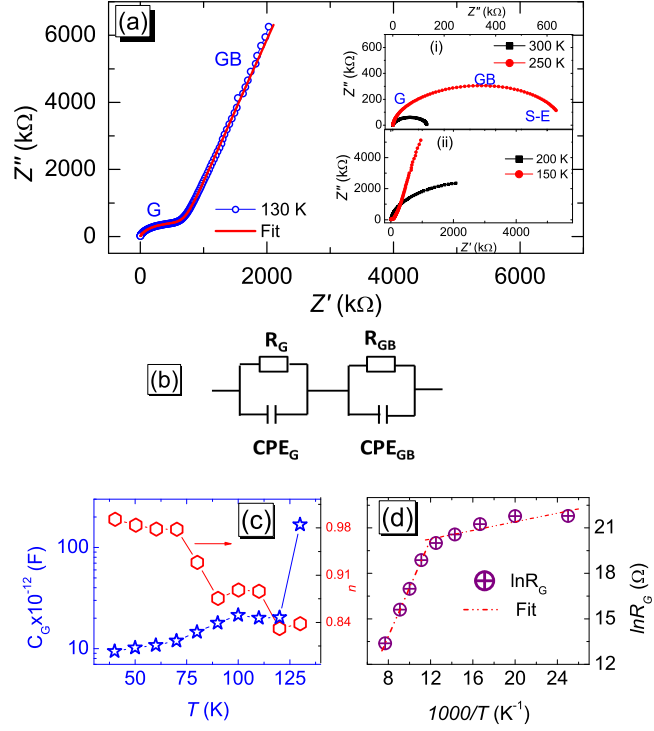}
	\caption{(colour online) (a) The $Z'$ versus $Z''$ plot at 130~K along with the equivalent circuit analysis fit;  $Z'$ versus $Z''$ plot at selected temperatures are given in insets (i) and (ii) of (a) where the contributions from grain (G), grain-boundary (GB) and sample-electrode (S-E) are marked. (b) Equivalent circuit. (c) Temperature dependence of bulk (grain) capacitance and critical exponent ($n$) from Eqn.~(\ref{imp_eqn}); both show anomalies near $T_c$. (d)Plot of ln($R_G$) $vs$ 1000/$T$ where the slope change near $T_c$ is evident. }
	\label{fig4_impedance}
\end{figure}
% #############################################################################################################################
%
\indent
It is relevant to confirm if the origin of experimentally observed colossal dielectric constant near room temperature is intrinsic or extrinsic. In polycrystalline perovskite oxides, a major contribution to the dielectric constant occurs from the grain boundaries.\cite{lunkenheimer2009colossal} The relaxation observed in such systems is due to electrical inhomogeneity of the sample where the charge carriers inside the grains are trapped by high-potential of the grain boundary regions, since grain boundaries act as very thin capacitors resulting in a large net capacitance.\cite{lunkenheimer2002origin} The space-charge relaxes at the grain boundaries leading to Maxwell-Wagner (MW) type of polarization. Electrical response of grain boundaries is associated with larger capacitance and resistance compared to the grains.\cite{lunkenheimer2002origin,lunkenheimer2009colossal} Generally, their response frequency is much lower than that of grains and, due to the high resistance, a strong peak  results in the measured impedance. 
To elucidate the origin of colossal dielectric constant in YNMO, we measured the complex impedance of the material using impedance spectroscopy. Each component of relaxation can be modeled by a combination of resistors ($R$) and capacitors ($C$). In the ideal case of single relaxation, impedance plot ($Z''$ vs $Z'$) will be a perfect semicircle. In the case of non-ideal relaxation, the ideal capacitor ($C$) is replaced by a constant phase element ($\mathrm{CPE}$). The complex impedance of $\mathrm{CPE}$ is defined as\cite{K.1983},
%
% EQUATION 2
\begin{equation}
Z^*_\mathrm{CPE} = \frac{1}{C_\mathrm{CPE}(i\omega)^n}
\label{imp_eqn}
\end{equation}
where, $C_\mathrm{CPE}$ is the $\mathrm{CPE}$-specific capacitance. $\omega$ is the angular frequency and $n$ is a critical exponent with typical values between 0.6 and 1 (for ideal capacitor, $n$ = 1). $\mathrm{CPE}$ capacitance can be converted into real capacitance using a standard procedure.\cite{Hsu2001} The impedance of YNMO sample was measured at different temperatures and the $Z''$ vs $Z'$ plots at few selected temperatures are shown in Fig.~\ref{fig4_impedance} (a). The room temperature $\epsilon'$ -- $f$ plot resulted in three plateaus with likely contributions of  three different relaxations (inset of Fig.\ref{fig3_dielectric}) arising from grain (G), grain-boundaries (GB) and the sample-electrode (S-E) interface.\cite{Schmidt2007,Nhalil2013} In general, it is reckoned that a low-frequency response originates from S-E, intermediate-frequency response from GB and high-frequency response from G. At room temperature, all three contributions are clearly seen as different plateaus. The $Z''$ -- $Z'$ plot at 300~K can be deconvoluted by applying three R-CPE units. Below 150~K, sample-interface contribution vanishes and the intrinsic bulk contribution starts to appear as a separate distorted semicircle in the high frequency region. Above this temperature, separating the bulk contribution from the GB and the interface components is a non-trivial task. The main panel of Fig.~\ref{fig4_impedance} (a) shows the plot at 130~K along with the equivalent circuit analysis fit using the circuit shown in the panel (b). Below 100~K, the GB contribution entirely disappears from the frequency window leaving only the bulk contribution which is then modeled by a single R-CPE circuit.
\\
\indent
Values of capacitance ($C_\mathrm{G}$), critical exponent ($n$) and resistance ($R_\mathrm{G}$) of the bulk are calculated from the equivalent circuit analysis and are shown in Fig.~\ref{fig4_impedance} (c) and (d). Bulk dielectric constant is calculated from $C_\mathrm{G}$ using the formula $\epsilon_r$ = $C_\mathrm{G}$/$\epsilon_0$. From Fig~\ref{fig4_impedance} (c), it is clear that $C_\mathrm{G}$ and $n$ reveal an anomaly at $T_c$. In the ln($R_\mathrm{G}$) -- 1000/$T$ plot, two regions with different slopes -- one above and the other below $T_c$ -- (Fig.~\ref{fig4_impedance} (d)) are observed, similar to that seen in the dielectric constant analysis. The activation energy ($E_a$) obtained here, above  and below the $T_c$, are 137.5(4)~meV and 12.2(1)~meV respectively. These values are comparable with those obtained earlier. It is noted that $E_a$ below $T_c$ is lower in magnitude than that above $T_c$ here as well.
\\
\subsection{Pyroelectric Current Measurements}
\label{pyro}
\begin{figure}[!t]
	\centering
	\includegraphics[scale=0.33]{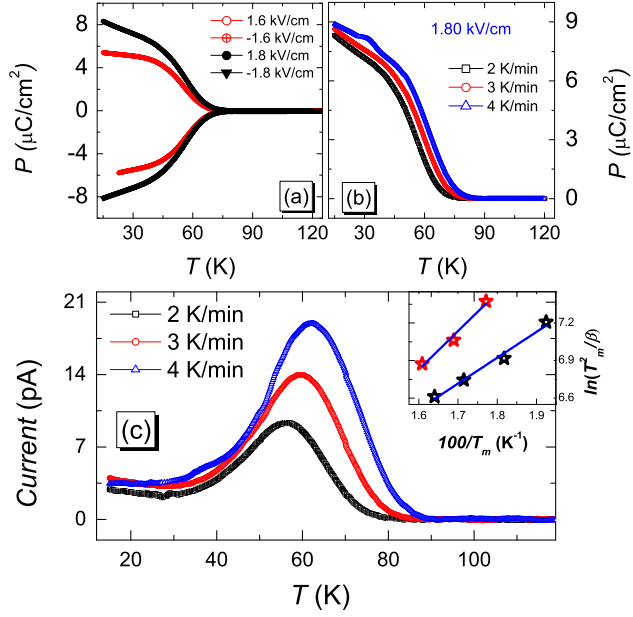}
	\caption{\label{fig5_pyro} (colour online) (a) Electrical polarization ($P$) as a function of temperature for different poling fields and poling directions. (b) Temperature dependence of polarization is shown for three different ramp rates. (c)  Pyrocurrent dispersion at three different ramp rates. The inset displays the ln($T^2_m$/$\beta$) vs 100/$T_m$ plot and the fit according to Eqn.~(\ref{pyro_eqn}) for two samples (open red stars-S1 and black S2).}
\end{figure}

\indent
Pyroelectric response of YNMO was measured at low temperature (below 120~K) in order to check whether the anomalies seen in the dielectric and impedance data are related to spontaneous electric polarization. The sample was poled at  high field ($\ge$ 1.5~kV/cm) well  above the $T_c$, cooled to 15~K and short-circuited for one hour to remove any accumulated space-charges. The current is registered while warming the dielectric at a constant ramp rate and spontaneous polarization is calculated by integrating the current with respect to time. Electric polarization is calculated from the measured pyrocurrents during forward and reverse poling at different poling fields. The experimentally measured polarization ($P$) is presented in Fig.~\ref{fig5_pyro} (a). Data is presented for positive and negative poling fields of $\pm$ 1.6~kV/cm and $\pm$ 1.8~kV/cm. Fig.~\ref{fig5_pyro} (b) confirms that the polarization as well as its saturation is dependent on ramp rate. The measured pyrocurrent at three different ramp rates are presented in Fig.~\ref{fig5_pyro} (c). A broad peak is visible centered at 57~K and reaching up to 62 K depending on the ramp rate. Dispersion in data with the ramp rate is clearly evident. To ensure that the observed property is intrinsic in nature, the same measurements were repeated on another sample (Sample 2, S2) of same batch where similar ramp rate dependent change in peak temperature was observed. Here, the peak temperature changed from 52 to 62 K when the ramp rate was varied from 2 to 5 K/min. The broad peak in pyrocurrent, slow dynamics of dipoles and ramp rate-dependent polarization are testimony to extrinsic origin of the pyrocurrent in YNMO. This encourages us to
%
%Figure 6 ##########################################################################################################################
% #######################################################################################################################################
%
%
postulate that {\em {thermally stimulated free charge carriers}} (TSFC) are responsible for the pyrocurrent peak in YNMO. The electric field poling procedure creates an internal electric field by the dipoles from frozen free charge carriers.\cite{de2015effect} The internal field  acts opposite to the applied poling field. Even after removal of applied field and shorting the sample for few hours, these dipoles persist. The relaxation time for such dipoles is in the order of several hours. When the dielectric is heated at a constant rate $\beta$ = $dT/dt$, the relaxation time of frozen dipoles will shorten. \cite{bucci1966ionic} This produces a depolarizing current as the dipoles gradually lose their preferred orientation. The current is likely to increase exponentially at first, reach a  threshold maximum and then gradually drops to zero.  In case of reverse poling, a similar freeze-and-release process is likely to occur but the current  flows in the reverse direction. Unless carefully monitored, current reversal can be misunderstood by its resemblance to ferroelectric polarization. From the peak position of the observed current, the activation energy and relaxation time can be calculated using the formula \cite{kohara2010excess,zhang2014investigation},
%
% EQUATION 3 ########################################################################################################################
\begin{equation}
\mathrm{ln}\left(\frac{T^2_m}{\beta}\right) = \left(\frac{E_a}{k_BT_m}\right) + \mathrm{ln}\left(\frac{\tau_0E_a}{k_B}\right)
\label{pyro_eqn}
\end{equation}  	   
% ###################################################################################################################################
%
where, $T_m$ is the temperature corresponding to the pyrocurrent peak, $\beta$ is the temperature ramp rate, $\tau_0$ is the relaxation time and $E_a$ is the activation energy. ln($T^2_m$/$\beta$) versus 100/$T_m$ is plotted in the inset of Fig.~\ref{fig5_pyro} (c). The activation energy calculated as per the fit is 260(2)~meV (176(3)~meV) and the relaxation time $\tau_0$ = 2.3(3) $\times$ 10$^{-3}$s (1.21(4) $\times$ 10$^{-2}$s) (values of S2  are given in bracket). %This value of activation energy can be compared with that calculated from dielectric relaxation which is ascribed to the charge transfer between Mn$^{4+} $ and Ni$^{2+}$.
The slow decay of current when the temperature ramping is stopped explains the relaxation times. It is relevant to note that the temperature dependent dielectric constant did not show any anomaly around this temperature ({57~K).
\\
%
%Figure 6 ##########################################################################################################################
	\begin{figure}[!t]
		\centering
		\includegraphics[scale=0.35]{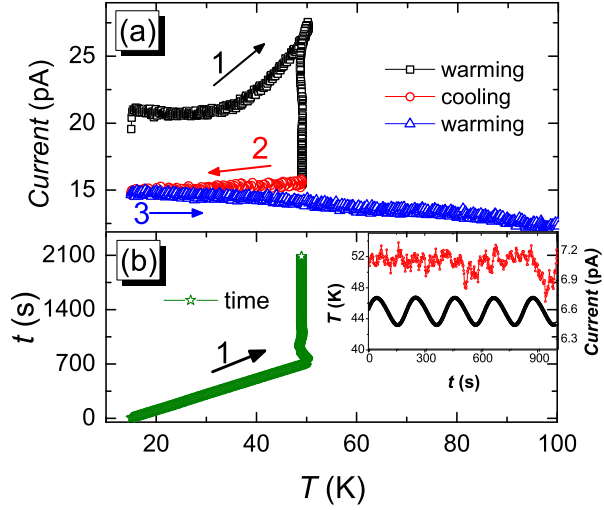}
		\caption{\label{fig6_pyro} (colour online) (a) Temperature dependence of pyrocurrent using special  measurement protocol. Measurement cycles are numbered in order and the arrows indicate the direction of heating/cooling cycles. (b) Slow decay after a halt at 50~K and the time elapsed corresponds to the time taken for the current to fall to the background value in the first cycle. Inset shows the temperature cycling result performed at 45 K. Detailed measurement protocols are given in the text.}
\end{figure}
% #######################################################################################################################################
	%
\indent 
In order to confirm the ferroelectric phase, a special protocol was used in which the sample was poled from a temperature well above the ferromagnetic $T_c$ as usual and the poling field was removed at 15~K. Subsequently, the sample was shorted for one hour and then heated up to 50 K (below 57~K, the pyrocurrent peak temperature) with a ramp rate of 3 K/min where the temperature ramping was stopped. For a ferroelectric, the pyrocurrent should immediately fall to zero (or to the background value) when the temperature is held constant since the relaxation dynamics last for few micro-seconds. However, we noted that the current drops very slowly taking several minutes ($\approx$35 min, Fig.~\ref{fig6_pyro} (b)) to reach the background value. Immediately after the current reaches  the background value, the sample was cooled  to 15 K with the same ramp rate as in the warming cycle ({\em i.e.,} 3 K/min). Since pyrocurrent is proportional to rate of change of temperature ($\beta$ = dT/dt), a sudden fall in current in the -ve direction is expected in the beginning of this cooling cycle and a gradual decrease thereafter as the temperature lowered. Reversal of current is expected because the sign of $\beta$ is reversed in the cooling cycle. In our measurements, we could not identify any current reversal. After reaching the temperature at 15 K, the sample is again heated, this time crossing the peak temperature, 57 K. Since we had never crossed the pyrocurrent peak temperature in the previous cycles, one would expect that the spontaneous polarization exists in the system. Spontaneous polarization will disappear only when the material crosses the transition (para to ferro) temperature. But, here, there was no sign of pyrocurrent (not even a small peak) in the third cycle thus ruling out the possibility of spontaneous polarization and ferroelectricity in our measurement. These results are presented in Figs.~\ref{fig6_pyro} (a) and (b). Even though the equivalent circuit analysis of dielectric and impedance studies (section~\ref{impedance}) revealed an anomaly near the FM $T_c$, intrinsic ferroelectric phase in YNMO is confirmed to be absent through this measurement. This result add credence to the proposed TSFC hypothesis of extrinsic polarization. When the system is poled and heated at constant ramp rate from 15 K, these charge carriers (TSFC) that are polarized and frozen-in begin to depolarize and a current is generated. It increases exponentially as the temperature rises and reaches a maximum value until the time of highest  rate of depolarization. In the first cycle, heating is stopped at  50 K, just below the peak temperature and is allowed to halt at this point until the magnitude of  current falls to the background value. As mentioned earlier, it was nearly 35 minutes before the background value was realized. Most carriers depolarize during the waiting period leaving only few polarized charge carriers. During the next cycle (cooling), all unpolarized charges along with the few remaining polarized charges freezes. Further, in the third cycle (heating), there are almost no dipoles to depolarize and negligible pyrocurrent is produced.\\
\indent
We measured the pyrocurrent by cycling  within a short temperature range below the peak temperature to confirm the extrinsic origin of the pyrocurrent. The temperature is sinusoidally varied centered at 45 K (45$\pm$1.5 K). It is known that the true pyrocurrent lags behind the temperature by 90 $^\circ$ and the current generated by TSFC will remain in-phase with the temperature.\cite{garn1982use,sharp1982use} The temperature cycling was performed for several hours to discharge all residual currents. In the beginning , the current followed the temperature in-phase with a gradual decrease in magnitude with time (fig not shown) but once the extrinsic contribution ceased, the current should follow the temperature with 90 $^\circ$ phase lag, if there is any spontaneous polarization. The measurement result in YNMO after the removal of the TSFC contribution is shown in the inset of Fig.~\ref{fig6_pyro} (b). No pyrocurrent systematics could be  observed.  This result establishes conclusively the extrinsic origin of pyroelectric currents in YNMO.
\section{Summary and Conclusions}
\label{discussion}
Using first-principles density functional theory calculations and model Hamiltonian analysis, it was possible to relate the changes of  the magnetic order in $R_2$NiMnO$_6$ compounds corresponding to FM ($R$ = La and Sm) and $E^*$-type (Y).\cite{kumar2010theoretical} In $E^*$-type, the magnetic structure consists of $\uparrow\uparrow\downarrow\downarrow$ spin chains along the cubic directions which breaks inversion symmetry and allows electric polarization to occur. It was also suggested by Kumar {\em et al.,}\cite{kumar2010theoretical} that the magnetic transition from FM to $E^*$-type can be effected by the application of electric field. Neutron powder diffraction experiments clearly reveal that the nuclear structure of YNMO has $P2_1/n$ symmetry with ordered arrangement of Ni$^{2+}$ and Mn$^{4+}$. However, the refined lattice parameters differ slightly from the values adapted for theoretical work.\cite{kumar2010theoretical} The experimentally estimated bond angle Ni--O--Mn lies close to the theoretical estimates. In accordance with the reported predictions, we observed the FM ground state of YNMO by refining the magnetic structure at 10~K. FM alignment of Ni$^{2+}$ and Mn$^{4+}$ is discernible from the saturation magnetization data at 5~K and 50~kOe. It is surmised that high degree of cationic ordering in the compound (favoured by Goodenough-Kanamori rules) results in ferromagnetism. In-plane Ni-Ni second nearest neighbour ($J^\mathrm{NNN}_{||\mathrm{(Ni-Ni)}}$), in-plane Mn-Mn second-nearest neighbour ($J^\mathrm{NNN}_{||\mathrm{(Mn-Mn)}}$), out-of-plane Ni-Mn nearest neighbour ($J^\mathrm{NN}_{\perp\mathrm{(Ni-Mn)}}$) and in-plane Ni-Mn nearest-neighbour ($J^\mathrm{NN}_{||\mathrm{(Ni-Mn)}}$) exchange constants reduce in magnitude and this trend is commonly observed in double perovskite compounds when $R$ is changed from La to Y.\cite{kumar2010theoretical} The ratio of $J^\mathrm{NNN}_{||\mathrm{(Ni-Ni)}}$ and $J^\mathrm{NN}_{\perp\mathrm{(Ni-Mn)}}$ increases during the transition to Y which stabilizes the $E^*$-type magnetic structure. One can find clear differences between the magnetic structure of YNMO and the multiferroic Lu$_2$CoMnO$_6$ \cite{yanez2011multiferroic} where magnetism-driven ferroelectricity is reported. In Lu$_2$CoMnO$_6$,  $\uparrow\uparrow\downarrow\downarrow$ magnetic structure is deduced without any spin canting while in YNMO and Lu$_2$NiMnO$_6$ \cite{manna2015structural} canted FM spin structure (canting with respect to crystallographic c axis) is observed. This canting may have an adverse effect in realizing magnetism-driven ferroelectricity, and could be the reason for the absence of ferroelectric phase in YNMO studied here. 
\\
\indent
A recent report on magnetism-driven multiferroic behavior of YNMO is in contradiction to our results.\cite{su2015magnetism} There are explicit differences between our material's magnetic and dielectric properties vis-a-vis those of reported.  The prominent M-H hysteresis loop had a coercive field of $\approx$0.27 T and saturation magnetization of 3.9 $\mu_B$/f.u. at 6 T field in the literature report whereas we do not see any prominent hysteresis while the saturation magnetization is 4.3(1)$\mu_B$/f.u. at 5 T field, higher than Su {\em et al} \cite{su2015magnetism}, nearer to the theoretical values for Mn$^{4+}$ -- Ni$^{2+}$ combination. Neutron studies confirm this and FM order in our material. This suggests that the magnetism in these two cases are slightly different. A dielectric anomaly at FM transition temperature was earlier reported \cite{su2015magnetism} which is absent in our studies. Instead, we see a broad hump in the vicinity of FM transition in dielectric constant data.  Ramp rate dependent dielectric studies revealed that this feature has similar characteristics as the one  observed in pyrocurrent measurements. The peak temperature increased with increase in ramp rate. Absence of ferroelectric phase is confirmed through detailed pyrocurrent measurements. However, a study of the equivalent circuit analysis of impedance data reveals the anomaly that coincides with the FM $T_c$ in the capacitance and resistance data. This would imply that  a relation exists between magnetism and dielectric properties of YNMO, although, it does not warrant a magneto-dielectric effect. While, pyrocurrent suggests the possibility of switchable polarization ( demonstrated here at $\pm$ 1.6 and $\pm$ 1.8~kV/cm), detailed measurements indicate that this is more likely to be  resulting from the {\em{thermally stimulated free charge carriers}}. Comparing with the magnetic structure of Lu$_2$CoMnO$_6$ where magnetism driven ferroelectricity is reported, a canted FM spin structure can be seen in YNMO. This could be the reason for the absence of ferroelectricity in YNMO.\\
\indent
In conclusion, Y$_2$NiMnO$_6$ is identified as a ferromagnet with  $T_c \approx$ 81~K. Crystal and magnetic structure analyzed by neutron diffraction data highlights a high degree of cationic ordering among Mn$^{4+}$ and Ni$^{2+}$ ions. Even though no dielectric anomalies were explicitly visible coinciding with  $T_c$ in the dielectric response, detailed equivalent circuit analysis of dielectric and impedance data showed correlation between electric and magnetic degrees of freedom in this compound. From pyrocurrent studies, it is concluded that the polarization in Y$_2$NiMnO$_6$ is deceptive and that the experimentally ascertained magnetic structure is in conformity with that of the theoretical predictions. The canted FM spin magnetic structure is hindering the magnetism-driven ferroelectricity in YNMO. 
\section*{Acknowledgments}
Thanks are due to Aditya Wagh and Ruchika Yadav for help with pyrocurrent measurements. One of us, (A.M.S.) thanks the SA-NRF (93549) and the FRC/URC
of UJ for financial assistance. H.S.N. acknowledges FRC/URC for a postdoctoral fellowship. Part of the research conducted at ORNL's Spallation Neutron Source was sponsored by the Scientific User Facilities Division, Office of Basic Energy Sciences, US Department of Energy. 
% 
%
% REFERENCES
%
\bibliographystyle{aipnum4-1}
%\bibliography{YNMO}

\begin{thebibliography}{49}%
	\makeatletter
	\providecommand \@ifxundefined [1]{%
		\@ifx{#1\undefined}
	}%
	\providecommand \@ifnum [1]{%
		\ifnum #1\expandafter \@firstoftwo
		\else \expandafter \@secondoftwo
		\fi
	}%
	\providecommand \@ifx [1]{%
		\ifx #1\expandafter \@firstoftwo
		\else \expandafter \@secondoftwo
		\fi
	}%
	\providecommand \natexlab [1]{#1}%
	\providecommand \enquote  [1]{``#1''}%
	\providecommand \bibnamefont  [1]{#1}%
	\providecommand \bibfnamefont [1]{#1}%
	\providecommand \citenamefont [1]{#1}%
	\providecommand \href@noop [0]{\@secondoftwo}%
	\providecommand \href [0]{\begingroup \@sanitize@url \@href}%
	\providecommand \@href[1]{\@@startlink{#1}\@@href}%
	\providecommand \@@href[1]{\endgroup#1\@@endlink}%
	\providecommand \@sanitize@url [0]{\catcode `\\12\catcode `\$12\catcode
		`\&12\catcode `\#12\catcode `\^12\catcode `\_12\catcode `\%12\relax}%
	\providecommand \@@startlink[1]{}%
	\providecommand \@@endlink[0]{}%
	\providecommand \url  [0]{\begingroup\@sanitize@url \@url }%
	\providecommand \@url [1]{\endgroup\@href {#1}{\urlprefix }}%
	\providecommand \urlprefix  [0]{URL }%
	\providecommand \Eprint [0]{\href }%
	\providecommand \doibase [0]{http://dx.doi.org/}%
	\providecommand \selectlanguage [0]{\@gobble}%
	\providecommand \bibinfo  [0]{\@secondoftwo}%
	\providecommand \bibfield  [0]{\@secondoftwo}%
	\providecommand \translation [1]{[#1]}%
	\providecommand \BibitemOpen [0]{}%
	\providecommand \bibitemStop [0]{}%
	\providecommand \bibitemNoStop [0]{.\EOS\space}%
	\providecommand \EOS [0]{\spacefactor3000\relax}%
	\providecommand \BibitemShut  [1]{\csname bibitem#1\endcsname}%
	\let\auto@bib@innerbib\@empty
	%</preamble>
	\bibitem [{\citenamefont {Rogado}\ \emph {et~al.}(2005)\citenamefont {Rogado},
		\citenamefont {Li}, \citenamefont {Sleight},\ and\ \citenamefont
		{Subramanian}}]{rogado2005magnetocapacitance}%
	\BibitemOpen
	\bibfield  {author} {\bibinfo {author} {\bibfnamefont {N.~S.}\ \bibnamefont
			{Rogado}}, \bibinfo {author} {\bibfnamefont {J.}~\bibnamefont {Li}}, \bibinfo
		{author} {\bibfnamefont {A.~W.}\ \bibnamefont {Sleight}}, \ and\ \bibinfo
		{author} {\bibfnamefont {M.~A.}\ \bibnamefont {Subramanian}},\ }\href@noop {}
	{\bibfield  {journal} {\bibinfo  {journal} {Adv. Mater.}\ }\textbf {\bibinfo
			{volume} {17}},\ \bibinfo {pages} {2225} (\bibinfo {year}
		{2005})}\BibitemShut {NoStop}%
	\bibitem [{\citenamefont {Nair.}\ \emph {et~al.}(2011)\citenamefont {Nair.},
		\citenamefont {Swain}, \citenamefont {Hariharan}, \citenamefont {Adiga},
		\citenamefont {Narayana},\ and\ \citenamefont
		{Elzabeth}}]{nair2011griffiths}%
	\BibitemOpen
	\bibfield  {author} {\bibinfo {author} {\bibfnamefont {H.~S.}\ \bibnamefont
			{Nair.}}, \bibinfo {author} {\bibfnamefont {D.}~\bibnamefont {Swain}},
		\bibinfo {author} {\bibfnamefont {N.}~\bibnamefont {Hariharan}}, \bibinfo
		{author} {\bibfnamefont {S.}~\bibnamefont {Adiga}}, \bibinfo {author}
		{\bibfnamefont {C.}~\bibnamefont {Narayana}}, \ and\ \bibinfo {author}
		{\bibfnamefont {S.}~\bibnamefont {Elzabeth}},\ }\href@noop {} {\bibfield
		{journal} {\bibinfo  {journal} {J. Appl. Phys.}\ }\textbf {\bibinfo {volume}
			{110}},\ \bibinfo {pages} {123919} (\bibinfo {year} {2011})}\BibitemShut
	{NoStop}%
	\bibitem [{\citenamefont {Guo}\ \emph {et~al.}(2013)\citenamefont {Guo},
		\citenamefont {Shi}, \citenamefont {Zhou}, \citenamefont {Zhao},\ and\
		\citenamefont {Liu}}]{guo2013near}%
	\BibitemOpen
	\bibfield  {author} {\bibinfo {author} {\bibfnamefont {Y.}~\bibnamefont
			{Guo}}, \bibinfo {author} {\bibfnamefont {L.}~\bibnamefont {Shi}}, \bibinfo
		{author} {\bibfnamefont {S.}~\bibnamefont {Zhou}}, \bibinfo {author}
		{\bibfnamefont {J.}~\bibnamefont {Zhao}}, \ and\ \bibinfo {author}
		{\bibfnamefont {W.}~\bibnamefont {Liu}},\ }\href@noop {} {\bibfield
		{journal} {\bibinfo  {journal} {Appl. Phys. Lett.}\ }\textbf {\bibinfo
			{volume} {102}},\ \bibinfo {pages} {222401} (\bibinfo {year}
		{2013})}\BibitemShut {NoStop}%
	\bibitem [{\citenamefont {Macedo~Filho}, \citenamefont {Ayala},\ and\
		\citenamefont {de~Araujo~Paschoal}(2013)}]{macedo2013spin}%
	\BibitemOpen
	\bibfield  {author} {\bibinfo {author} {\bibfnamefont {R.~B.}\ \bibnamefont
			{Macedo~Filho}}, \bibinfo {author} {\bibfnamefont {A.~P.}\ \bibnamefont
			{Ayala}}, \ and\ \bibinfo {author} {\bibfnamefont {C.~W.}\ \bibnamefont
			{de~Araujo~Paschoal}},\ }\href@noop {} {\bibfield  {journal} {\bibinfo
			{journal} {Appl. Phys. Lett.}\ }\textbf {\bibinfo {volume} {102}},\ \bibinfo
		{pages} {192902} (\bibinfo {year} {2013})}\BibitemShut {NoStop}%
	\bibitem [{\citenamefont {Anderson}\ \emph {et~al.}(1993)\citenamefont
		{Anderson}, \citenamefont {Greenwood}, \citenamefont {Taylor},\ and\
		\citenamefont {Poeppelmeier}}]{anderson1993b}%
	\BibitemOpen
	\bibfield  {author} {\bibinfo {author} {\bibfnamefont {M.~T.}\ \bibnamefont
			{Anderson}}, \bibinfo {author} {\bibfnamefont {K.~B.}\ \bibnamefont
			{Greenwood}}, \bibinfo {author} {\bibfnamefont {G.~A.}\ \bibnamefont
			{Taylor}}, \ and\ \bibinfo {author} {\bibfnamefont {K.~R.}\ \bibnamefont
			{Poeppelmeier}},\ }\href@noop {} {\bibfield  {journal} {\bibinfo  {journal}
			{Prog. Solid State Chem.}\ }\textbf {\bibinfo {volume} {22}},\ \bibinfo
		{pages} {197} (\bibinfo {year} {1993})}\BibitemShut {NoStop}%
	\bibitem [{\citenamefont {Goodenough}(1955)}]{goodenough1955theory}%
	\BibitemOpen
	\bibfield  {author} {\bibinfo {author} {\bibfnamefont {J.~B.}\ \bibnamefont
			{Goodenough}},\ }\href@noop {} {\bibfield  {journal} {\bibinfo  {journal}
			{Phys. Rev.}\ }\textbf {\bibinfo {volume} {100}},\ \bibinfo {pages} {564}
		(\bibinfo {year} {1955})}\BibitemShut {NoStop}%
	\bibitem [{\citenamefont {Zhou}\ \emph {et~al.}(2010)\citenamefont {Zhou},
		\citenamefont {Guo}, \citenamefont {Zhao}, \citenamefont {Zhao},\ and\
		\citenamefont {Shi}}]{zhou2010nature}%
	\BibitemOpen
	\bibfield  {author} {\bibinfo {author} {\bibfnamefont {S.~M.}\ \bibnamefont
			{Zhou}}, \bibinfo {author} {\bibfnamefont {Y.~Q.}\ \bibnamefont {Guo}},
		\bibinfo {author} {\bibfnamefont {J.~Y.}\ \bibnamefont {Zhao}}, \bibinfo
		{author} {\bibfnamefont {S.~Y.}\ \bibnamefont {Zhao}}, \ and\ \bibinfo
		{author} {\bibfnamefont {L.}~\bibnamefont {Shi}},\ }\href@noop {} {\bibfield
		{journal} {\bibinfo  {journal} {Appl. Phys. Lett.}\ }\textbf {\bibinfo
			{volume} {96}},\ \bibinfo {pages} {262507} (\bibinfo {year}
		{2010})}\BibitemShut {NoStop}%
	\bibitem [{\citenamefont {Kumar}\ \emph {et~al.}(2010)\citenamefont {Kumar},
		\citenamefont {Giovannetti}, \citenamefont {van~den Brink},\ and\
		\citenamefont {Picozzi}}]{kumar2010theoretical}%
	\BibitemOpen
	\bibfield  {author} {\bibinfo {author} {\bibfnamefont {S.}~\bibnamefont
			{Kumar}}, \bibinfo {author} {\bibfnamefont {G.}~\bibnamefont {Giovannetti}},
		\bibinfo {author} {\bibfnamefont {J.}~\bibnamefont {van~den Brink}}, \ and\
		\bibinfo {author} {\bibfnamefont {S.}~\bibnamefont {Picozzi}},\ }\href@noop
	{} {\bibfield  {journal} {\bibinfo  {journal} {Phys. Rev. B}\ }\textbf
		{\bibinfo {volume} {82}},\ \bibinfo {pages} {134429} (\bibinfo {year}
		{2010})}\BibitemShut {NoStop}%
	\bibitem [{\citenamefont {Y{\'a}{\~n}ez-Vilar}\ \emph
		{et~al.}(2011)\citenamefont {Y{\'a}{\~n}ez-Vilar}, \citenamefont {Mun},
		\citenamefont {Zapf}, \citenamefont {Ueland}, \citenamefont {Gardner},
		\citenamefont {Thompson}, \citenamefont {Singleton}, \citenamefont
		{S{\'a}nchez-And{\'u}jar}, \citenamefont {Mira}, \citenamefont {Biskup} \emph
		{et~al.}}]{yanez2011multiferroic}%
	\BibitemOpen
	\bibfield  {author} {\bibinfo {author} {\bibfnamefont {S.}~\bibnamefont
			{Y{\'a}{\~n}ez-Vilar}}, \bibinfo {author} {\bibfnamefont {E.~D.}\
			\bibnamefont {Mun}}, \bibinfo {author} {\bibfnamefont {V.~S.}\ \bibnamefont
			{Zapf}}, \bibinfo {author} {\bibfnamefont {B.~G.}\ \bibnamefont {Ueland}},
		\bibinfo {author} {\bibfnamefont {J.~S.}\ \bibnamefont {Gardner}}, \bibinfo
		{author} {\bibfnamefont {J.~D.}\ \bibnamefont {Thompson}}, \bibinfo {author}
		{\bibfnamefont {J.}~\bibnamefont {Singleton}}, \bibinfo {author}
		{\bibfnamefont {M.}~\bibnamefont {S{\'a}nchez-And{\'u}jar}}, \bibinfo
		{author} {\bibfnamefont {J.}~\bibnamefont {Mira}}, \bibinfo {author}
		{\bibfnamefont {N.}~\bibnamefont {Biskup}},  \emph {et~al.},\ }\href@noop {}
	{\bibfield  {journal} {\bibinfo  {journal} {Phys. Rev. B}\ }\textbf {\bibinfo
			{volume} {84}},\ \bibinfo {pages} {134427} (\bibinfo {year}
		{2011})}\BibitemShut {NoStop}%
	\bibitem [{\citenamefont {Sharma}\ \emph {et~al.}(2013)\citenamefont {Sharma},
		\citenamefont {Saha}, \citenamefont {Kaushik}, \citenamefont {Siruguri},\
		and\ \citenamefont {Patnaik}}]{sharma2013magnetism}%
	\BibitemOpen
	\bibfield  {author} {\bibinfo {author} {\bibfnamefont {G.}~\bibnamefont
			{Sharma}}, \bibinfo {author} {\bibfnamefont {J.}~\bibnamefont {Saha}},
		\bibinfo {author} {\bibfnamefont {S.~D.}\ \bibnamefont {Kaushik}}, \bibinfo
		{author} {\bibfnamefont {V.}~\bibnamefont {Siruguri}}, \ and\ \bibinfo
		{author} {\bibfnamefont {S.}~\bibnamefont {Patnaik}},\ }\href@noop {}
	{\bibfield  {journal} {\bibinfo  {journal} {Appl. Phys. Lett.}\ }\textbf
		{\bibinfo {volume} {103}},\ \bibinfo {pages} {012903} (\bibinfo {year}
		{2013})}\BibitemShut {NoStop}%
	\bibitem [{\citenamefont {Su}\ \emph {et~al.}(2015)\citenamefont {Su},
		\citenamefont {Yang}, \citenamefont {Lu}, \citenamefont {Zhang},
		\citenamefont {Gu}, \citenamefont {Lu}, \citenamefont {Li}, \citenamefont
		{Liu},\ and\ \citenamefont {Zhu}}]{su2015magnetism}%
	\BibitemOpen
	\bibfield  {author} {\bibinfo {author} {\bibfnamefont {J.}~\bibnamefont
			{Su}}, \bibinfo {author} {\bibfnamefont {Z.}~\bibnamefont {Yang}}, \bibinfo
		{author} {\bibfnamefont {X.}~\bibnamefont {Lu}}, \bibinfo {author}
		{\bibfnamefont {J.}~\bibnamefont {Zhang}}, \bibinfo {author} {\bibfnamefont
			{L.}~\bibnamefont {Gu}}, \bibinfo {author} {\bibfnamefont {C.}~\bibnamefont
			{Lu}}, \bibinfo {author} {\bibfnamefont {Q.}~\bibnamefont {Li}}, \bibinfo
		{author} {\bibfnamefont {J.}~\bibnamefont {Liu}}, \ and\ \bibinfo {author}
		{\bibfnamefont {J.}~\bibnamefont {Zhu}},\ }\href@noop {} {\bibfield
		{journal} {\bibinfo  {journal} {ACS Appl. Mater. Inter.}\ } (\bibinfo {year}
		{2015})}\BibitemShut {NoStop}%
	\bibitem [{\citenamefont {Kakarla}\ \emph {et~al.}(2014)\citenamefont
		{Kakarla}, \citenamefont {Jyothinagaram}, \citenamefont {Das},\ and\
		\citenamefont {Adyam}}]{kakarla2014dielectric}%
	\BibitemOpen
	\bibfield  {author} {\bibinfo {author} {\bibfnamefont {D.~C.}\ \bibnamefont
			{Kakarla}}, \bibinfo {author} {\bibfnamefont {K.~M.}\ \bibnamefont
			{Jyothinagaram}}, \bibinfo {author} {\bibfnamefont {A.~K.}\ \bibnamefont
			{Das}}, \ and\ \bibinfo {author} {\bibfnamefont {V.}~\bibnamefont {Adyam}},\
	}\href@noop {} {\bibfield  {journal} {\bibinfo  {journal} {J. Am. Ceram.
			Soc.}\ }\textbf {\bibinfo {volume} {97}},\ \bibinfo {pages} {2858} (\bibinfo
	{year} {2014})}\BibitemShut {NoStop}%
\bibitem [{\citenamefont {Mouallem-Bahout}\ \emph {et~al.}(2004)\citenamefont
	{Mouallem-Bahout}, \citenamefont {Roisnel}, \citenamefont {Andr{\'e}},
	\citenamefont {Gutierrez}, \citenamefont {Moure},\ and\ \citenamefont
	{Pe{\~n}a}}]{mouallem2004nuclear}%
\BibitemOpen
\bibfield  {author} {\bibinfo {author} {\bibfnamefont {M.}~\bibnamefont
		{Mouallem-Bahout}}, \bibinfo {author} {\bibfnamefont {T.}~\bibnamefont
		{Roisnel}}, \bibinfo {author} {\bibfnamefont {G.}~\bibnamefont {Andr{\'e}}},
	\bibinfo {author} {\bibfnamefont {D.}~\bibnamefont {Gutierrez}}, \bibinfo
	{author} {\bibfnamefont {C.}~\bibnamefont {Moure}}, \ and\ \bibinfo {author}
	{\bibfnamefont {O.}~\bibnamefont {Pe{\~n}a}},\ }\href@noop {} {\bibfield
	{journal} {\bibinfo  {journal} {Solid State Commun.}\ }\textbf {\bibinfo
		{volume} {129}},\ \bibinfo {pages} {255} (\bibinfo {year}
	{2004})}\BibitemShut {NoStop}%
\bibitem [{\citenamefont {Schmidt}\ \emph {et~al.}(2007)\citenamefont
	{Schmidt}, \citenamefont {Eerenstein}, \citenamefont {Winiecki},
	\citenamefont {Morrison},\ and\ \citenamefont {Midgley}}]{Schmidt2007}%
\BibitemOpen
\bibfield  {author} {\bibinfo {author} {\bibfnamefont {R.}~\bibnamefont
		{Schmidt}}, \bibinfo {author} {\bibfnamefont {W.}~\bibnamefont {Eerenstein}},
	\bibinfo {author} {\bibfnamefont {T.}~\bibnamefont {Winiecki}}, \bibinfo
	{author} {\bibfnamefont {F.~D.}\ \bibnamefont {Morrison}}, \ and\ \bibinfo
	{author} {\bibfnamefont {P.~A.}\ \bibnamefont {Midgley}},\ }\href@noop {}
{\bibfield  {journal} {\bibinfo  {journal} {Phy. Rev. B}\ }\textbf {\bibinfo
		{volume} {245111}},\ \bibinfo {pages} {1} (\bibinfo {year}
	{2007})}\BibitemShut {NoStop}%
\bibitem [{\citenamefont {Schmidt}\ \emph {et~al.}(2012)\citenamefont
	{Schmidt}, \citenamefont {Ventura}, \citenamefont {Langenberg}, \citenamefont
	{Nemes}, \citenamefont {Munuera}, \citenamefont {Varela}, \citenamefont
	{Garcia-hernandez}, \citenamefont {Leon}, \citenamefont {Santamaria},\ and\
	\citenamefont {Kelvin}}]{Schmidt2012}%
\BibitemOpen
\bibfield  {author} {\bibinfo {author} {\bibfnamefont {R.}~\bibnamefont
		{Schmidt}}, \bibinfo {author} {\bibfnamefont {J.}~\bibnamefont {Ventura}},
	\bibinfo {author} {\bibfnamefont {E.}~\bibnamefont {Langenberg}}, \bibinfo
	{author} {\bibfnamefont {N.~M.}\ \bibnamefont {Nemes}}, \bibinfo {author}
	{\bibfnamefont {C.}~\bibnamefont {Munuera}}, \bibinfo {author} {\bibfnamefont
		{M.}~\bibnamefont {Varela}}, \bibinfo {author} {\bibfnamefont
		{M.}~\bibnamefont {Garcia-hernandez}}, \bibinfo {author} {\bibfnamefont
		{C.}~\bibnamefont {Leon}}, \bibinfo {author} {\bibfnamefont {J.}~\bibnamefont
		{Santamaria}}, \ and\ \bibinfo {author} {\bibfnamefont {T.}~\bibnamefont
		{Kelvin}},\ }\href@noop {} {\bibfield  {journal} {\bibinfo  {journal} {Phys.
			Rev. B}\ }\textbf {\bibinfo {volume} {035113}},\ \bibinfo {pages} {1}
	(\bibinfo {year} {2012})}\BibitemShut {NoStop}%
\bibitem [{\citenamefont {Schmidt}\ \emph {et~al.}(2009)\citenamefont
	{Schmidt}, \citenamefont {Wu}, \citenamefont {Leighton},\ and\ \citenamefont
	{Terry}}]{schmidt2009dielectric}%
\BibitemOpen
\bibfield  {author} {\bibinfo {author} {\bibfnamefont {R.}~\bibnamefont
		{Schmidt}}, \bibinfo {author} {\bibfnamefont {J.}~\bibnamefont {Wu}},
	\bibinfo {author} {\bibfnamefont {C.}~\bibnamefont {Leighton}}, \ and\
	\bibinfo {author} {\bibfnamefont {I.}~\bibnamefont {Terry}},\ }\href@noop {}
{\bibfield  {journal} {\bibinfo  {journal} {Phys. Rev. B}\ }\textbf {\bibinfo
		{volume} {79}},\ \bibinfo {pages} {125105} (\bibinfo {year}
	{2009})}\BibitemShut {NoStop}%
\bibitem [{\citenamefont {Nair}\ \emph {et~al.}(2014)\citenamefont {Nair},
	\citenamefont {Pradheesh}, \citenamefont {Xiao}, \citenamefont {Cherian},
	\citenamefont {Elizabeth}, \citenamefont {Hansen}, \citenamefont
	{Chatterji},\ and\ \citenamefont {Br{\"u}ckel}}]{nair2014magnetizationsteps}%
\BibitemOpen
\bibfield  {author} {\bibinfo {author} {\bibfnamefont {H.~S.}\ \bibnamefont
		{Nair}}, \bibinfo {author} {\bibfnamefont {R.}~\bibnamefont {Pradheesh}},
	\bibinfo {author} {\bibfnamefont {Y.}~\bibnamefont {Xiao}}, \bibinfo {author}
	{\bibfnamefont {D.}~\bibnamefont {Cherian}}, \bibinfo {author} {\bibfnamefont
		{S.}~\bibnamefont {Elizabeth}}, \bibinfo {author} {\bibfnamefont
		{T.}~\bibnamefont {Hansen}}, \bibinfo {author} {\bibfnamefont
		{T.}~\bibnamefont {Chatterji}}, \ and\ \bibinfo {author} {\bibfnamefont
		{T.}~\bibnamefont {Br{\"u}ckel}},\ }\href@noop {} {\bibfield  {journal}
	{\bibinfo  {journal} {J. Appl. Phys.}\ }\textbf {\bibinfo {volume} {116}},\
	\bibinfo {pages} {123907} (\bibinfo {year} {2014})}\BibitemShut {NoStop}%
\bibitem [{\citenamefont {Huq}\ \emph {et~al.}(2011)\citenamefont {Huq},
	\citenamefont {Hodges}, \citenamefont {Gourdon},\ and\ \citenamefont
	{Heroux}}]{AHuq_1ZK2011AThirdGeneration}%
\BibitemOpen
\bibfield  {author} {\bibinfo {author} {\bibfnamefont {A.}~\bibnamefont
		{Huq}}, \bibinfo {author} {\bibfnamefont {J.~P.}\ \bibnamefont {Hodges}},
	\bibinfo {author} {\bibfnamefont {O.}~\bibnamefont {Gourdon}}, \ and\
	\bibinfo {author} {\bibfnamefont {L.}~\bibnamefont {Heroux}},\ }\href@noop {}
{\bibfield  {journal} {\bibinfo  {journal} {Z. Kristallogr. Proc.}\ }\textbf
	{\bibinfo {volume} {1}},\ \bibinfo {pages} {127} (\bibinfo {year}
	{2011})}\BibitemShut {NoStop}%
\bibitem [{\citenamefont {Rietveld}(1969)}]{rietveld}%
\BibitemOpen
\bibfield  {author} {\bibinfo {author} {\bibfnamefont {H.~M.}\ \bibnamefont
		{Rietveld}},\ }\href@noop {} {\bibfield  {journal} {\bibinfo  {journal} {J.
			Appl. Cryst.}\ }\textbf {\bibinfo {volume} {2}},\ \bibinfo {pages} {65}
	(\bibinfo {year} {1969})}\BibitemShut {NoStop}%
\bibitem [{\citenamefont {Rodriguez-Carvajal}(2010)}]{fullprof}%
\BibitemOpen
\bibfield  {author} {\bibinfo {author} {\bibfnamefont {J.}~\bibnamefont
		{Rodriguez-Carvajal}},\ }\href@noop {} {\bibfield  {journal} {\bibinfo
		{journal} {LLB, CEA-CNRS, France [http://www. ill. eu/sites/fullprof/]}\ }
	(\bibinfo {year} {2010})}\BibitemShut {NoStop}%
\bibitem [{\citenamefont {Maiti}\ \emph {et~al.}(2012)\citenamefont {Maiti},
	\citenamefont {Dutta}, \citenamefont {Mukherjee}, \citenamefont {Mitra},\
	and\ \citenamefont {Chakravorty}}]{maiti2012magnetic}%
\BibitemOpen
\bibfield  {author} {\bibinfo {author} {\bibfnamefont {R.~P.}\ \bibnamefont
		{Maiti}}, \bibinfo {author} {\bibfnamefont {S.}~\bibnamefont {Dutta}},
	\bibinfo {author} {\bibfnamefont {M.}~\bibnamefont {Mukherjee}}, \bibinfo
	{author} {\bibfnamefont {M.~K.}\ \bibnamefont {Mitra}}, \ and\ \bibinfo
	{author} {\bibfnamefont {D.}~\bibnamefont {Chakravorty}},\ }\href@noop {}
{\bibfield  {journal} {\bibinfo  {journal} {J. Appl. Phys.}\ }\textbf
	{\bibinfo {volume} {112}},\ \bibinfo {pages} {044311} (\bibinfo {year}
	{2012})}\BibitemShut {NoStop}%
\bibitem [{\citenamefont {Asaka}\ \emph {et~al.}(2007)\citenamefont {Asaka},
	\citenamefont {Yu}, \citenamefont {Tomioka}, \citenamefont {Kaneko},
	\citenamefont {Nagai}, \citenamefont {Kimoto}, \citenamefont {Ishizuka},
	\citenamefont {Tokura},\ and\ \citenamefont {Matsui}}]{asaka2007strong}%
\BibitemOpen
\bibfield  {author} {\bibinfo {author} {\bibfnamefont {T.}~\bibnamefont
		{Asaka}}, \bibinfo {author} {\bibfnamefont {X.~Z.}\ \bibnamefont {Yu}},
	\bibinfo {author} {\bibfnamefont {Y.}~\bibnamefont {Tomioka}}, \bibinfo
	{author} {\bibfnamefont {Y.}~\bibnamefont {Kaneko}}, \bibinfo {author}
	{\bibfnamefont {T.}~\bibnamefont {Nagai}}, \bibinfo {author} {\bibfnamefont
		{K.}~\bibnamefont {Kimoto}}, \bibinfo {author} {\bibfnamefont
		{K.}~\bibnamefont {Ishizuka}}, \bibinfo {author} {\bibfnamefont
		{Y.}~\bibnamefont {Tokura}}, \ and\ \bibinfo {author} {\bibfnamefont
		{Y.}~\bibnamefont {Matsui}},\ }\href@noop {} {\bibfield  {journal} {\bibinfo
		{journal} {Phys. Rev. B}\ }\textbf {\bibinfo {volume} {75}},\ \bibinfo
	{pages} {184440} (\bibinfo {year} {2007})}\BibitemShut {NoStop}%
\bibitem [{\citenamefont {Dass}, \citenamefont {Yan},\ and\ \citenamefont
	{Goodenough}(2003)}]{dass2003oxygen}%
\BibitemOpen
\bibfield  {author} {\bibinfo {author} {\bibfnamefont {R.~I.}\ \bibnamefont
		{Dass}}, \bibinfo {author} {\bibfnamefont {J.-Q.}\ \bibnamefont {Yan}}, \
	and\ \bibinfo {author} {\bibfnamefont {J.~B.}\ \bibnamefont {Goodenough}},\
}\href@noop {} {\bibfield  {journal} {\bibinfo  {journal} {Phys. Rev. B}\
}\textbf {\bibinfo {volume} {68}},\ \bibinfo {pages} {064415} (\bibinfo
{year} {2003})}\BibitemShut {NoStop}%
\bibitem [{\citenamefont {Ogale}\ \emph {et~al.}(1999)\citenamefont {Ogale},
	\citenamefont {Ogale}, \citenamefont {Ramesh},\ and\ \citenamefont
	{Venkatesan}}]{ogale1999octahedral}%
\BibitemOpen
\bibfield  {author} {\bibinfo {author} {\bibfnamefont {A.~S.}\ \bibnamefont
		{Ogale}}, \bibinfo {author} {\bibfnamefont {S.~B.}\ \bibnamefont {Ogale}},
	\bibinfo {author} {\bibfnamefont {R.}~\bibnamefont {Ramesh}}, \ and\ \bibinfo
	{author} {\bibfnamefont {T.}~\bibnamefont {Venkatesan}},\ }\href@noop {}
{\bibfield  {journal} {\bibinfo  {journal} {Appl. Phy. Lett.}\ }\textbf
	{\bibinfo {volume} {75}},\ \bibinfo {pages} {537} (\bibinfo {year}
	{1999})}\BibitemShut {NoStop}%
\bibitem [{\citenamefont {Dagotto}, \citenamefont {Hotta},\ and\ \citenamefont
	{Moreo}(2001)}]{dagotto2001colossal}%
\BibitemOpen
\bibfield  {author} {\bibinfo {author} {\bibfnamefont {E.}~\bibnamefont
		{Dagotto}}, \bibinfo {author} {\bibfnamefont {T.}~\bibnamefont {Hotta}}, \
	and\ \bibinfo {author} {\bibfnamefont {A.}~\bibnamefont {Moreo}},\
}\href@noop {} {\bibfield  {journal} {\bibinfo  {journal} {Phys. Rep.}\
}\textbf {\bibinfo {volume} {344}},\ \bibinfo {pages} {1} (\bibinfo {year}
{2001})}\BibitemShut {NoStop}%
\bibitem [{\citenamefont {Li}\ \emph {et~al.}(2005)\citenamefont {Li},
	\citenamefont {Shen}, \citenamefont {Thompson},\ and\ \citenamefont
	{Weitering}}]{li2005ferromagnetic}%
\BibitemOpen
\bibfield  {author} {\bibinfo {author} {\bibfnamefont {A.~P.}\ \bibnamefont
		{Li}}, \bibinfo {author} {\bibfnamefont {J.}~\bibnamefont {Shen}}, \bibinfo
	{author} {\bibfnamefont {J.~R.}\ \bibnamefont {Thompson}}, \ and\ \bibinfo
	{author} {\bibfnamefont {H.~H.}\ \bibnamefont {Weitering}},\ }\href@noop {}
{\bibfield  {journal} {\bibinfo  {journal} {Appl. Phys. Lett.}\ }\textbf
	{\bibinfo {volume} {86}},\ \bibinfo {pages} {152507} (\bibinfo {year}
	{2005})}\BibitemShut {NoStop}%
\bibitem [{\citenamefont {He}\ \emph {et~al.}(2007)\citenamefont {He},
	\citenamefont {Torija}, \citenamefont {Wu}, \citenamefont {Lynn},
	\citenamefont {Zheng}, \citenamefont {Mitchell},\ and\ \citenamefont
	{Leighton}}]{he2007non}%
\BibitemOpen
\bibfield  {author} {\bibinfo {author} {\bibfnamefont {C.}~\bibnamefont
		{He}}, \bibinfo {author} {\bibfnamefont {M.~A.}\ \bibnamefont {Torija}},
	\bibinfo {author} {\bibfnamefont {J.}~\bibnamefont {Wu}}, \bibinfo {author}
	{\bibfnamefont {J.~W.}\ \bibnamefont {Lynn}}, \bibinfo {author}
	{\bibfnamefont {H.}~\bibnamefont {Zheng}}, \bibinfo {author} {\bibfnamefont
		{J.~F.}\ \bibnamefont {Mitchell}}, \ and\ \bibinfo {author} {\bibfnamefont
		{C.}~\bibnamefont {Leighton}},\ }\href@noop {} {\bibfield  {journal}
	{\bibinfo  {journal} {Phys. Rev. B}\ }\textbf {\bibinfo {volume} {76}},\
	\bibinfo {pages} {014401} (\bibinfo {year} {2007})}\BibitemShut {NoStop}%
\bibitem [{\citenamefont {Zhou}\ \emph {et~al.}(2011)\citenamefont {Zhou},
	\citenamefont {Guo}, \citenamefont {Zhao}, \citenamefont {He},\ and\
	\citenamefont {Shi}}]{zhou2011size}%
\BibitemOpen
\bibfield  {author} {\bibinfo {author} {\bibfnamefont {S.}~\bibnamefont
		{Zhou}}, \bibinfo {author} {\bibfnamefont {Y.}~\bibnamefont {Guo}}, \bibinfo
	{author} {\bibfnamefont {J.}~\bibnamefont {Zhao}}, \bibinfo {author}
	{\bibfnamefont {L.}~\bibnamefont {He}}, \ and\ \bibinfo {author}
	{\bibfnamefont {L.}~\bibnamefont {Shi}},\ }\href@noop {} {\bibfield
	{journal} {\bibinfo  {journal} {J. Phys. Chem. C}\ }\textbf {\bibinfo
		{volume} {115}},\ \bibinfo {pages} {1535} (\bibinfo {year}
	{2011})}\BibitemShut {NoStop}%
\bibitem [{\citenamefont {Saber}\ \emph {et~al.}(2010)\citenamefont {Saber},
	\citenamefont {Egilmez}, \citenamefont {Mansour}, \citenamefont {Fan},
	\citenamefont {Chow},\ and\ \citenamefont {Jung}}]{saber2010evolution}%
\BibitemOpen
\bibfield  {author} {\bibinfo {author} {\bibfnamefont {M.~M.}\ \bibnamefont
		{Saber}}, \bibinfo {author} {\bibfnamefont {M.}~\bibnamefont {Egilmez}},
	\bibinfo {author} {\bibfnamefont {A.~I.}\ \bibnamefont {Mansour}}, \bibinfo
	{author} {\bibfnamefont {I.}~\bibnamefont {Fan}}, \bibinfo {author}
	{\bibfnamefont {K.~H.}\ \bibnamefont {Chow}}, \ and\ \bibinfo {author}
	{\bibfnamefont {J.}~\bibnamefont {Jung}},\ }\href@noop {} {\bibfield
	{journal} {\bibinfo  {journal} {Phys. Rev. B}\ }\textbf {\bibinfo {volume}
		{82}},\ \bibinfo {pages} {172401} (\bibinfo {year} {2010})}\BibitemShut
{NoStop}%
\bibitem [{\citenamefont {Biswal}\ \emph {et~al.}(2014)\citenamefont {Biswal},
	\citenamefont {Ray}, \citenamefont {Babu}, \citenamefont {Siruguri},\ and\
	\citenamefont {Vishwakarma}}]{biswal_jap_115_2014dielectric}%
\BibitemOpen
\bibfield  {author} {\bibinfo {author} {\bibfnamefont {A.~K.}\ \bibnamefont
		{Biswal}}, \bibinfo {author} {\bibfnamefont {J.}~\bibnamefont {Ray}},
	\bibinfo {author} {\bibfnamefont {P.~D.}\ \bibnamefont {Babu}}, \bibinfo
	{author} {\bibfnamefont {V.}~\bibnamefont {Siruguri}}, \ and\ \bibinfo
	{author} {\bibfnamefont {P.~N.}\ \bibnamefont {Vishwakarma}},\ }\href@noop {}
{\bibfield  {journal} {\bibinfo  {journal} {J. Appl. Phys.}\ }\textbf
	{\bibinfo {volume} {115}},\ \bibinfo {pages} {194106} (\bibinfo {year}
	{2014})}\BibitemShut {NoStop}%
\bibitem [{\citenamefont {Liu}\ \emph {et~al.}(2014)\citenamefont {Liu},
	\citenamefont {Shi}, \citenamefont {Zhou}, \citenamefont {Zhao},
	\citenamefont {Li},\ and\ \citenamefont {Guo}}]{liu_jap_116_2014griffiths}%
\BibitemOpen
\bibfield  {author} {\bibinfo {author} {\bibfnamefont {W.}~\bibnamefont
		{Liu}}, \bibinfo {author} {\bibfnamefont {L.}~\bibnamefont {Shi}}, \bibinfo
	{author} {\bibfnamefont {S.}~\bibnamefont {Zhou}}, \bibinfo {author}
	{\bibfnamefont {J.}~\bibnamefont {Zhao}}, \bibinfo {author} {\bibfnamefont
		{Y.}~\bibnamefont {Li}}, \ and\ \bibinfo {author} {\bibfnamefont
		{Y.}~\bibnamefont {Guo}},\ }\href@noop {} {\bibfield  {journal} {\bibinfo
		{journal} {J. Appl. Phys.}\ }\textbf {\bibinfo {volume} {116}},\ \bibinfo
	{pages} {193901} (\bibinfo {year} {2014})}\BibitemShut {NoStop}%
\bibitem [{\citenamefont {Bull}, \citenamefont {Gleeson},\ and\ \citenamefont
	{Knight}(2003)}]{bulljpcm_15_4927_2003}%
\BibitemOpen
\bibfield  {author} {\bibinfo {author} {\bibfnamefont {C.~L.}\ \bibnamefont
		{Bull}}, \bibinfo {author} {\bibfnamefont {D.}~\bibnamefont {Gleeson}}, \
	and\ \bibinfo {author} {\bibfnamefont {K.~S.}\ \bibnamefont {Knight}},\
}\href@noop {} {\bibfield  {journal} {\bibinfo  {journal} {J. Phys.: Condens.
		Matter}\ }\textbf {\bibinfo {volume} {15}},\ \bibinfo {pages} {4927}
(\bibinfo {year} {2003})}\BibitemShut {NoStop}%
\bibitem [{\citenamefont {Wills}(2000)}]{wills_sarah}%
\BibitemOpen
\bibfield  {author} {\bibinfo {author} {\bibfnamefont {A.~S.}\ \bibnamefont
		{Wills}},\ }\href@noop {} {\bibfield  {journal} {\bibinfo  {journal} {Physica
			B}\ }\textbf {\bibinfo {volume} {276}},\ \bibinfo {pages} {680} (\bibinfo
	{year} {2000})}\BibitemShut {NoStop}%
\bibitem [{\citenamefont {Zhang}\ \emph
	{et~al.}(2014{\natexlab{a}})\citenamefont {Zhang}, \citenamefont {Zhang},
	\citenamefont {Ge}, \citenamefont {Wang}, \citenamefont {Yuan},\ and\
	\citenamefont {Feng}}]{zhang2014hydrothermal}%
\BibitemOpen
\bibfield  {author} {\bibinfo {author} {\bibfnamefont {C.}~\bibnamefont
		{Zhang}}, \bibinfo {author} {\bibfnamefont {T.}~\bibnamefont {Zhang}},
	\bibinfo {author} {\bibfnamefont {L.}~\bibnamefont {Ge}}, \bibinfo {author}
	{\bibfnamefont {S.}~\bibnamefont {Wang}}, \bibinfo {author} {\bibfnamefont
		{H.}~\bibnamefont {Yuan}}, \ and\ \bibinfo {author} {\bibfnamefont
		{S.}~\bibnamefont {Feng}},\ }\href@noop {} {\bibfield  {journal} {\bibinfo
		{journal} {RSC Adv.}\ }\textbf {\bibinfo {volume} {4}},\ \bibinfo {pages}
	{50969} (\bibinfo {year} {2014}{\natexlab{a}})}\BibitemShut {NoStop}%
\bibitem [{\citenamefont {Cole}\ and\ \citenamefont {Cole}(1941)}]{Cole1941}%
\BibitemOpen
\bibfield  {author} {\bibinfo {author} {\bibfnamefont {K.~S.}\ \bibnamefont
		{Cole}}\ and\ \bibinfo {author} {\bibfnamefont {R.~H.}\ \bibnamefont
		{Cole}},\ }\href@noop {} {\bibfield  {journal} {\bibinfo  {journal} {J. Chem.
			Phys.}\ }\textbf {\bibinfo {volume} {9}},\ \bibinfo {pages} {341} (\bibinfo
	{year} {1941})}\BibitemShut {NoStop}%
\bibitem [{\citenamefont {Manna}\ \emph {et~al.}(2014)\citenamefont {Manna},
	\citenamefont {Joshi}, \citenamefont {Elizabeth},\ and\ \citenamefont
	{Anil~Kumar}}]{Manna2014}%
\BibitemOpen
\bibfield  {author} {\bibinfo {author} {\bibfnamefont {K.}~\bibnamefont
		{Manna}}, \bibinfo {author} {\bibfnamefont {R.~S.}\ \bibnamefont {Joshi}},
	\bibinfo {author} {\bibfnamefont {S.}~\bibnamefont {Elizabeth}}, \ and\
	\bibinfo {author} {\bibfnamefont {P.~S.}\ \bibnamefont {Anil~Kumar}},\
}\href@noop {} {\bibfield  {journal} {\bibinfo  {journal} {Appl. Phys.
		Lett.}\ }\textbf {\bibinfo {volume} {104}},\ \bibinfo {pages} {202905}
(\bibinfo {year} {2014})}\BibitemShut {NoStop}%
\bibitem [{\citenamefont {Yang}\ \emph {et~al.}(2012)\citenamefont {Yang},
	\citenamefont {He}, \citenamefont {Zhu}, \citenamefont {Bai}, \citenamefont
	{Sun}, \citenamefont {Meng}, \citenamefont {Tang}, \citenamefont {Duan},
	\citenamefont {Remiens}, \citenamefont {Qiu} \emph {et~al.}}]{yang2012small}%
\BibitemOpen
\bibfield  {author} {\bibinfo {author} {\bibfnamefont {J.}~\bibnamefont
		{Yang}}, \bibinfo {author} {\bibfnamefont {J.}~\bibnamefont {He}}, \bibinfo
	{author} {\bibfnamefont {J.~Y.}\ \bibnamefont {Zhu}}, \bibinfo {author}
	{\bibfnamefont {W.}~\bibnamefont {Bai}}, \bibinfo {author} {\bibfnamefont
		{L.}~\bibnamefont {Sun}}, \bibinfo {author} {\bibfnamefont {X.~J.}\
		\bibnamefont {Meng}}, \bibinfo {author} {\bibfnamefont {X.~D.}\ \bibnamefont
		{Tang}}, \bibinfo {author} {\bibfnamefont {C.-G.}\ \bibnamefont {Duan}},
	\bibinfo {author} {\bibfnamefont {D.}~\bibnamefont {Remiens}}, \bibinfo
	{author} {\bibfnamefont {J.~H.}\ \bibnamefont {Qiu}},  \emph {et~al.},\
}\href@noop {} {\bibfield  {journal} {\bibinfo  {journal} {Appl. Phys.
		Lett.}\ }\textbf {\bibinfo {volume} {101}},\ \bibinfo {pages} {222904}
(\bibinfo {year} {2012})}\BibitemShut {NoStop}%
\bibitem [{\citenamefont {Lunkenheimer}\ \emph {et~al.}(2009)\citenamefont
	{Lunkenheimer}, \citenamefont {Krohns}, \citenamefont {Riegg}, \citenamefont
	{Ebbinghaus}, \citenamefont {Reller},\ and\ \citenamefont
	{Loidl}}]{lunkenheimer2009colossal}%
\BibitemOpen
\bibfield  {author} {\bibinfo {author} {\bibfnamefont {P.}~\bibnamefont
		{Lunkenheimer}}, \bibinfo {author} {\bibfnamefont {S.}~\bibnamefont
		{Krohns}}, \bibinfo {author} {\bibfnamefont {S.}~\bibnamefont {Riegg}},
	\bibinfo {author} {\bibfnamefont {S.~G.}\ \bibnamefont {Ebbinghaus}},
	\bibinfo {author} {\bibfnamefont {A.}~\bibnamefont {Reller}}, \ and\ \bibinfo
	{author} {\bibfnamefont {A.}~\bibnamefont {Loidl}},\ }\href@noop {}
{\bibfield  {journal} {\bibinfo  {journal} {Eur. Phys. J. Spec. Top.}\
	}\textbf {\bibinfo {volume} {180}},\ \bibinfo {pages} {61} (\bibinfo {year}
	{2009})}\BibitemShut {NoStop}%
\bibitem [{\citenamefont {Lunkenheimer}\ \emph {et~al.}(2002)\citenamefont
	{Lunkenheimer}, \citenamefont {Bobnar}, \citenamefont {Pronin}, \citenamefont
	{Ritus}, \citenamefont {Volkov},\ and\ \citenamefont
	{Loidl}}]{lunkenheimer2002origin}%
\BibitemOpen
\bibfield  {author} {\bibinfo {author} {\bibfnamefont {P.}~\bibnamefont
		{Lunkenheimer}}, \bibinfo {author} {\bibfnamefont {V.}~\bibnamefont
		{Bobnar}}, \bibinfo {author} {\bibfnamefont {A.~V.}\ \bibnamefont {Pronin}},
	\bibinfo {author} {\bibfnamefont {A.~I.}\ \bibnamefont {Ritus}}, \bibinfo
	{author} {\bibfnamefont {A.~A.}\ \bibnamefont {Volkov}}, \ and\ \bibinfo
	{author} {\bibfnamefont {A.}~\bibnamefont {Loidl}},\ }\href@noop {}
{\bibfield  {journal} {\bibinfo  {journal} {Phys. Rev. B}\ }\textbf {\bibinfo
		{volume} {66}},\ \bibinfo {pages} {052105} (\bibinfo {year}
	{2002})}\BibitemShut {NoStop}%
\bibitem [{\citenamefont {Jonscher}(1983)}]{K.1983}%
\BibitemOpen
\bibfield  {author} {\bibinfo {author} {\bibfnamefont {A.~K.}\ \bibnamefont
		{Jonscher}},\ }\href@noop {} {\emph {\bibinfo {title} {Dielectric Relaxation
			in Solids}}}\ (\bibinfo  {publisher} {Chelsea Dielectric Press, London},\
\bibinfo {year} {1983})\BibitemShut {NoStop}%
\bibitem [{\citenamefont {Hsu}\ and\ \citenamefont {Mansfeld}(2011)}]{Hsu2001}%
\BibitemOpen
\bibfield  {author} {\bibinfo {author} {\bibfnamefont {C.~H.}\ \bibnamefont
		{Hsu}}\ and\ \bibinfo {author} {\bibfnamefont {F.}~\bibnamefont {Mansfeld}},\
}\href@noop {} {\bibfield  {journal} {\bibinfo  {journal} {Corrosios}\
}\textbf {\bibinfo {volume} {57}},\ \bibinfo {pages} {747} (\bibinfo {year}
{2011})}\BibitemShut {NoStop}%
\bibitem [{\citenamefont {Nhalil}\ \emph {et~al.}(2013)\citenamefont {Nhalil},
	\citenamefont {Nair}, \citenamefont {Bhat},\ and\ \citenamefont
	{Elizabeth}}]{Nhalil2013}%
\BibitemOpen
\bibfield  {author} {\bibinfo {author} {\bibfnamefont {H.}~\bibnamefont
		{Nhalil}}, \bibinfo {author} {\bibfnamefont {H.~S.}\ \bibnamefont {Nair}},
	\bibinfo {author} {\bibfnamefont {H.~L.}\ \bibnamefont {Bhat}}, \ and\
	\bibinfo {author} {\bibfnamefont {S.}~\bibnamefont {Elizabeth}},\ }\href@noop
{} {\bibfield  {journal} {\bibinfo  {journal} {Eur. Phys. Lett}\ }\textbf
	{\bibinfo {volume} {104}},\ \bibinfo {pages} {67002} (\bibinfo {year}
	{2013})}\BibitemShut {NoStop}%
\bibitem [{\citenamefont {De}, \citenamefont {Ghara},\ and\ \citenamefont
	{Sundaresan}(2015)}]{de2015effect}%
\BibitemOpen
\bibfield  {author} {\bibinfo {author} {\bibfnamefont {C.}~\bibnamefont
		{De}}, \bibinfo {author} {\bibfnamefont {S.}~\bibnamefont {Ghara}}, \ and\
	\bibinfo {author} {\bibfnamefont {A.}~\bibnamefont {Sundaresan}},\
}\href@noop {} {\bibfield  {journal} {\bibinfo  {journal} {Solid State
		Commun.}\ } (\bibinfo {year} {2015})}\BibitemShut {NoStop}%
\bibitem [{\citenamefont {Bucci}, \citenamefont {Fieschi},\ and\ \citenamefont
	{Guidi}(1966)}]{bucci1966ionic}%
\BibitemOpen
\bibfield  {author} {\bibinfo {author} {\bibfnamefont {C.}~\bibnamefont
		{Bucci}}, \bibinfo {author} {\bibfnamefont {R.}~\bibnamefont {Fieschi}}, \
	and\ \bibinfo {author} {\bibfnamefont {G.}~\bibnamefont {Guidi}},\
}\href@noop {} {\bibfield  {journal} {\bibinfo  {journal} {Phys. Rev.}\
}\textbf {\bibinfo {volume} {148}},\ \bibinfo {pages} {816} (\bibinfo {year}
{1966})}\BibitemShut {NoStop}%
\bibitem [{\citenamefont {Kohara}\ \emph {et~al.}(2010)\citenamefont {Kohara},
	\citenamefont {Yamasaki}, \citenamefont {Onose},\ and\ \citenamefont
	{Tokura}}]{kohara2010excess}%
\BibitemOpen
\bibfield  {author} {\bibinfo {author} {\bibfnamefont {Y.}~\bibnamefont
		{Kohara}}, \bibinfo {author} {\bibfnamefont {Y.}~\bibnamefont {Yamasaki}},
	\bibinfo {author} {\bibfnamefont {Y.}~\bibnamefont {Onose}}, \ and\ \bibinfo
	{author} {\bibfnamefont {Y.}~\bibnamefont {Tokura}},\ }\href@noop {}
{\bibfield  {journal} {\bibinfo  {journal} {Phys. Rev. B}\ }\textbf {\bibinfo
		{volume} {82}},\ \bibinfo {pages} {104419} (\bibinfo {year}
	{2010})}\BibitemShut {NoStop}%
\bibitem [{\citenamefont {Zhang}\ \emph
	{et~al.}(2014{\natexlab{b}})\citenamefont {Zhang}, \citenamefont {Zhao},
	\citenamefont {Cui}, \citenamefont {Ye}, \citenamefont {Zhao}, \citenamefont
	{Li}, \citenamefont {Wang}, \citenamefont {Zhu}, \citenamefont {Zhang},\ and\
	\citenamefont {Rao}}]{zhang2014investigation}%
\BibitemOpen
\bibfield  {author} {\bibinfo {author} {\bibfnamefont {X.}~\bibnamefont
		{Zhang}}, \bibinfo {author} {\bibfnamefont {Y.~G.}\ \bibnamefont {Zhao}},
	\bibinfo {author} {\bibfnamefont {Y.~F.}\ \bibnamefont {Cui}}, \bibinfo
	{author} {\bibfnamefont {L.~D.}\ \bibnamefont {Ye}}, \bibinfo {author}
	{\bibfnamefont {D.~Y.}\ \bibnamefont {Zhao}}, \bibinfo {author}
	{\bibfnamefont {P.~S.}\ \bibnamefont {Li}}, \bibinfo {author} {\bibfnamefont
		{J.~W.}\ \bibnamefont {Wang}}, \bibinfo {author} {\bibfnamefont {M.~H.}\
		\bibnamefont {Zhu}}, \bibinfo {author} {\bibfnamefont {H.~Y.}\ \bibnamefont
		{Zhang}}, \ and\ \bibinfo {author} {\bibfnamefont {G.~H.}\ \bibnamefont
		{Rao}},\ }\href@noop {} {\bibfield  {journal} {\bibinfo  {journal} {Appl.
			Phys. Lett.}\ }\textbf {\bibinfo {volume} {104}},\ \bibinfo {pages} {062903}
	(\bibinfo {year} {2014}{\natexlab{b}})}\BibitemShut {NoStop}%
\bibitem [{\citenamefont {Garn}\ and\ \citenamefont
	{Sharp}(1982)}]{garn1982use}%
\BibitemOpen
\bibfield  {author} {\bibinfo {author} {\bibfnamefont {L.~E.}\ \bibnamefont
		{Garn}}\ and\ \bibinfo {author} {\bibfnamefont {E.~J.}\ \bibnamefont
		{Sharp}},\ }\href@noop {} {\bibfield  {journal} {\bibinfo  {journal} {J.
			Appl. Phys.}\ }\textbf {\bibinfo {volume} {53}},\ \bibinfo {pages} {8974}
	(\bibinfo {year} {1982})}\BibitemShut {NoStop}%
\bibitem [{\citenamefont {Sharp}\ and\ \citenamefont
	{Garn}(1982)}]{sharp1982use}%
\BibitemOpen
\bibfield  {author} {\bibinfo {author} {\bibfnamefont {E.~J.}\ \bibnamefont
		{Sharp}}\ and\ \bibinfo {author} {\bibfnamefont {L.~E.}\ \bibnamefont
		{Garn}},\ }\href@noop {} {\bibfield  {journal} {\bibinfo  {journal} {J. Appl.
			Phys.}\ }\textbf {\bibinfo {volume} {53}},\ \bibinfo {pages} {8980} (\bibinfo
	{year} {1982})}\BibitemShut {NoStop}%
\bibitem [{\citenamefont {Manna}\ \emph {et~al.}(2015)\citenamefont {Manna},
	\citenamefont {Bera}, \citenamefont {Jain}, \citenamefont {Elizabeth},
	\citenamefont {Yusuf},\ and\ \citenamefont {Kumar}}]{manna2015structural}%
\BibitemOpen
\bibfield  {author} {\bibinfo {author} {\bibfnamefont {K.}~\bibnamefont
		{Manna}}, \bibinfo {author} {\bibfnamefont {A.~K.}\ \bibnamefont {Bera}},
	\bibinfo {author} {\bibfnamefont {M.}~\bibnamefont {Jain}}, \bibinfo {author}
	{\bibfnamefont {S.}~\bibnamefont {Elizabeth}}, \bibinfo {author}
	{\bibfnamefont {S.~M.}\ \bibnamefont {Yusuf}}, \ and\ \bibinfo {author}
	{\bibfnamefont {P.~S.~A.}\ \bibnamefont {Kumar}},\ }\href@noop {} {\bibfield
	{journal} {\bibinfo  {journal} {Phys. Rev. B}\ }\textbf {\bibinfo {volume}
		{91}},\ \bibinfo {pages} {224420} (\bibinfo {year} {2015})}\BibitemShut
{NoStop}%
\end{thebibliography}
%

%
%
%
\end{document}